\documentclass[fleqn,usenatbib]{mnras}

\usepackage{newtxtext,newtxmath}

\usepackage[T1]{fontenc}
\usepackage{scalerel}
\usepackage{bbold}
\DeclareRobustCommand{\VAN}[3]{#2}
\let\VANthebibliography\thebibliography
\def\thebibliography{\DeclareRobustCommand{\VAN}[3]{##3}\VANthebibliography}

\usepackage{graphicx}	
\usepackage{amsmath}	

\title[Radial velocities via STFT]{A Linearized Approach to Radial-Velocity Extraction. II: Shot-Noise-Limited Precision via Spectral Factorization}

\author[S. Shahaf \& B. Zackay]{
S. Shahaf$^{1,2}$\thanks{E-mail: sashahaf@mpia.de}
and B. Zackay$^{2}$
\\
$^{1}${Max-Planck Institute for Astronomy, K\"onigstuhl 17, D-69117 Heidelberg, Germany}\\
$^{2}$Department of Particle Physics and Astrophysics, Weizmann Institute of Science, Rehovot 7610001, Israel\\
}

\date{Accepted XXX. Received YYY; in original form ZZZ}

\pubyear{2025}

\begin{document}
\label{firstpage}
\pagerange{\pageref{firstpage}--\pageref{lastpage}}
\maketitle

\begin{abstract}
We generalize the short-time Fourier transform (STFT) formalism for radial velocity extraction to cases where the underlying spectral components are unknown. The method factorizes a spectroscopic time series into principal spectra and time-dependent kernels, enabling simultaneous recovery of both. In Fourier space, each inverse-wavelength slice is decomposed by singular value decomposition, and radial velocity shifts are inferred from phase differences between epochs. In the high-SNR regime, this provides a linearized and statistically tractable estimate of differential velocities. The method is validated on synthetic and SOAP simulations and applied to EXPRES observations of HD~34411 and $\tau$~Ceti, recovering coherent signals and reaching the instrumental precision limit of ${\sim}30$~cm\,s\textsuperscript{-1}. Apart from p-mode modulation, the residuals show no significant long-term correlations and allow the detection of signals with semi-amplitudes down to ${\sim}50$~cm\,s\textsuperscript{-1} with $\lesssim10$~cm\,s\textsuperscript{-1} uncertainty. The framework thus enables extreme-precision radial velocity measurements in the presence of spectral variability, representing a step toward detecting and characterizing Earth-like planets around solar-type stars.
\end{abstract}

\begin{keywords}
exoplanets -- techniques: radial velocities -- methods: data analysis 
\end{keywords}



\section{Introduction}
\label{sec: introduction}
\defcitealias{Shahaf2023}{Paper I}

Extracting precise stellar radial velocities has become a central challenge in exoplanetary astronomy. With the advancement of modern spectrographs, the limiting factor is no longer instrumental stability but variability of the stellar spectra. Magnetic activity, convection, and surface inhomogeneities can mimic or obscure Doppler shifts, producing spurious periodicities or concealing genuine planetary signals \citep[e.g.,][and references therein]{cameron21, Zhao_L_2023, hara23, Burt2025}. 

This recognition has driven the development of techniques aimed at separating Doppler shifts from other activity-induced parasitic signals. Several broad, partially overlapping strategies have emerged. For instance, many pipelines use spectral compression, summarizing the information from thousands of lines, or multiple observing epochs, into a representative profile or a few informative quantities \citep[e.g.,][]{Zhao_L_2024, Klein2025, Salzer2025}. Others adopt per-line or per-region analyses, estimating velocities independently from different spectral lines or orders before combining them, and often utilize chromatic diagnostics to mitigate the impact of variability on the inferred velocities \citep[e.g.,][]{cretignier20, AlMoulla2024, Cretignier2024}. A third direction employs data-driven decompositions, projecting the time series onto low-dimensional variability components learned directly from the spectra \citep[e.g.,][]{cameron21, cretignier22}. All approaches are supported by detailed instrumental characterization \citep[e.g,][]{Cretignier2023}, forward spectral modeling \citep[e.g][]{palumbo22,  zhao23, Zhao_Y_2024}, and time-domain post-processing frameworks \citep[e.g.][]{Rescigno2025}.

In our previous work \citep[][hereafter Paper I]{Shahaf2023}, we introduced a complementary approach to this problem. Conceptually, it lies at the intersection of several of the broad categories discussed above. The key idea is that a stellar spectrum can be linearized and factorized into a small number of manageable components. Each component separates into a \textit{principal spectrum}, a chromatic function representing the rest-frame spectral response to a specific type of variability, and a \textit{principal kernel}, whose weights describe how that component contributes at the observer's frame. In Paper I, we assumed that the principal spectra were known and developed a method to derive the principal kernels by analyzing partially overlapping spectral segments in the Fourier domain. The resulting representation forms, in a sense, an idealized compression scheme, as we showed that it retains the full information content of the spectrum and constitutes a sufficient statistic. \textit{This manuscript extends Paper I to a case where the principal spectra are unknown.} 

As before, we consider a stellar spectrum $S(\lambda)$ sampled over a logarithmically uniform wavelength grid, where the difference between two consecutive points corresponds to a constant shift in velocity. This approach allows us to systematically model the emission from various positions on the star’s surface, with each position contributing uniquely to the observed spectrum due to the star’s rotation and other factors.
We began by describing the local emission profile, $f(\lambda_{\rm e}; \mathbfit{r})$, at a particular position on the star, denoted by the vector $\mathbfit{r}$. This profile is expressed as a linear combination of multiple principal spectra $f^{(k)}$, each weighted by a corresponding function $u_{k}$ that depends on the position $\mathbfit{r}$. This approach allows for decomposing complex stellar emission patterns into manageable components, 
\begin{equation}
    \label{eq: f rest frame}
    f(\lambda_{\rm e}; \, \mathbfit{r}) = 
    u_{_0}({\mathbfit{r}}) \, f^{{\scaleto{(0)}{5pt}}}(\lambda_{\rm e}) + 
    u_{_1}({\mathbfit{r}}) \, f^{{\scaleto{(1)}{5pt}}}(\lambda_{\rm e}) +  
    u_{_2}({\mathbfit{r}}) \, f^{{\scaleto{(2)}{5pt}}}(\lambda_{\rm e}) + \ldots \, ,
\end{equation}
where \(\lambda_{\rm e}\) is the emitted wavelength at the rest frame of the area element. The total observed spectrum \( S(\lambda) \) is then obtained by integrating these contributions over the entire visible disk of the star after shifting them to the observer's frame. This led to an expression for the disk-integrated spectrum as a sum of convolutions,
\begin{equation}
    S(\lambda) = U^{{\scaleto{(0)}{5pt}}} \circledast f^{{\scaleto{(0)}{5pt}}}(\lambda) + U^{{\scaleto{(1)}{5pt}}} \circledast f^{{\scaleto{(1)}{5pt}}}(\lambda) + \ldots \, ,
\end{equation}
where $U^{(k)}$ are referred to as the principal kernels and $f^{(k)}$ are the principal spectra mentioned above.

To simplify the inference procedure, we transitioned to the Fourier domain. Then, we utilized the chromatic behavior of the expansion by performing a Fourier transform on a small spectrum segment centered around some wavelength $\Lambda$:
\begin{equation}
    \tilde{S}_{\Lambda}(\zeta) = 
    \tilde{U}^{{\scaleto{(0)}{5pt}}}\, \tilde{f}^{{\scaleto{(0)}{5pt}}}_\Lambda(\zeta) + 
    \tilde{U}^{{\scaleto{(1)}{5pt}}}\, \tilde{f}^{{\scaleto{(1)}{5pt}}}_\Lambda(\zeta) + \ldots 
    \,.
    \label{eq: S Lambda}
\end{equation}
Here, $\tilde{f}^{{\scaleto{(k)}{5pt}}}_\Lambda$ represents the Fourier transform of the selected segment of the principal spectra, and the inverse-wavelength, $\zeta$, is the independent variable of the transform (i.e., conjugate to $\lambda$). This procedure essentially describes a short-time Fourier transform (STFT; e.g., \citealt{grochenig01}, \citealt{muller21}, \citealt{faulhuber22}).
Importantly, while the principal spectra change for different wavelength segments, their corresponding kernels remain unchanged. This is because the principal kernels, by construction, are insensitive to the specific wavelength domain. This property greatly simplifies the inference problem: by slicing the spectrum into several segments, we gain sufficient constraints to fit the kernel's Fourier coefficients since all segments share the same kernels. The readers are referred to \citetalias{Shahaf2023} for a detailed derivation and the equations. 

Our previous work required the chromatic response to variability to be known in advance. In many practical cases this response is not available and the principal spectra are therefore unknown. When a target is observed repeatedly, the time series itself can provide the information needed to infer these spectra. This work extends the previous formalism to this more general regime. Instead of assuming the principal spectra, we derive them empirically by factorizing a spectroscopic time series. The variability across epochs supplies the constraints needed to recover both the principal spectra and their associated kernels.
Section~\ref{sec: SVD} shows how a time series of spectroscopic measurements can be decomposed to derive the principal spectra and kernels simultaneously. Section~\ref{sec: phase retrieval} formulates the radial velocity extraction scheme, and demonstrates the process with two synthetic examples. Section~\ref{sec: test cases} validates the performance using spectroscopic data of HD~34411 and $\tau$~Ceti. Finally, in Section~\ref{sec: summary}, we summarize and discuss our results.


\section{Factorization of a time series}
\label{sec: SVD}
\subsection{Bilinear representation}
Consider a time series of spectroscopic measurements. As before, the principal spectra represent the chromatic components comprising the local emission profile. The coefficients scaling the contribution of each component now become a function of both the position on the star's luminous disk and the measurement time. Thus, equation~(\ref{eq: f rest frame}) can be rewritten as
\begin{equation}
    \label{eq: f rest frame time}
    f(\lambda_{\rm e}; \, \mathbfit{r}, t) = 
    u_{_0}({\mathbfit{r}}, t) \, f^{{\scaleto{(0)}{5pt}}}(\lambda_{\rm e}) + 
    u_{_1}({\mathbfit{r}}, t) \, f^{{\scaleto{(1)}{5pt}}}(\lambda_{\rm e})  + \ldots \, ,
\end{equation}
where $t$ represents the measurement time. Following the derivation of equation~(\ref{eq: S Lambda}), the Fourier transform of the spectral segment centered around $\Lambda$ becomes
\begin{equation}
    \tilde{S}_{t \Lambda}(\zeta) = 
    \tilde{U}^{{\scaleto{(0)}{5pt}}}_t\, \tilde{f}^{{\scaleto{(0)}{5pt}}}_\Lambda(\zeta) + 
    \tilde{U}^{{\scaleto{(1)}{5pt}}}_t\, \tilde{f}^{{\scaleto{(1)}{5pt}}}_\Lambda(\zeta) + \ldots \, .
    \label{eq: S t Lambda}
\end{equation}

Per spectrum, equations~(\ref{eq: S Lambda}) and~(\ref{eq: S t Lambda}) are identical. However, the latter equation representation expresses the principal kernels' role in encapsulating the spectral shape's time evolution. The equation above implies that $\tilde{S}_{t \Lambda}$ is a \textit{bilinear form}, namely, it can be expressed as a dot product of two vectors, 
\begin{equation}
    \label{eq: S bilinear}
     \tilde{S}_{t \Lambda}(\zeta) = \begin{pmatrix} \tilde{U}^{{\scaleto{(0)}{5pt}}}_t & \dots & \tilde{U}^{{\scaleto{(k)}{5pt}}}_t \end{pmatrix} \cdot
    \begin{pmatrix} \tilde{f}^{{\scaleto{(0)}{5pt}}}_\Lambda \\ \vdots \\ \tilde{f}^{{\scaleto{(k)}{5pt}}}_\Lambda\end{pmatrix}\,.
\end{equation}
This description will be used to decompose the entire spectroscopic time series.

\subsection{Singular value decomposition}
Per $\zeta$, the factorized spectroscopic time series can be expressed as a matrix, $\tilde{\mathbf{S}}_\zeta$, with rows corresponding to time-indexed data and columns representing spectral segments. This matrix has dimensions $n \times m$, where $n$ is the number of measurements, and $m$ denotes the spectral segments. 

As any other matrix, $\tilde{\mathbf{S}}_\zeta$ can be factorized using singular value decomposition (SVD), resulting in two unitary matrices of dimensions $n \times n$ and $m \times m$, respectively denoted $\mathbf{U}_\zeta$ and $\mathbf{V}_\zeta$. These unitary matrices contain the (left and right) singular vectors, while a third $n \times m$ diagonal matrix, denoted $\mathbf{\Sigma}_\zeta$, contains their corresponding real-valued non-negative singular values. The factorization is written as 
\begin{equation}
\label{eq: S SVD}
\tilde{\mathbf{S}}_\zeta = \mathbf{U}_\zeta \mathbf{\Sigma}_\zeta \mathbf{V}^{\dagger}_\zeta. 
\end{equation}
Singular value decomposition is presented and discussed in numerous textbooks. For example, see \citet[Chapter~2.6]{Horn2012}.

Equation~(\ref{eq: S bilinear}) implies that $\mathbf{U}_\zeta$ and $\mathbf{V}_\zeta^\dagger$ correspond to the principal kernels and spectra
at a given time and wavelength segment. This is evident from dimensional considerations: $n$ entries should be needed to describe the time-varying principal kernels; similarly, $m$ entries are required to describe the segmented principal spectra. The rank of the matrix, determined by the number of significant singular values in $\mathbf{\Sigma}_\zeta$, reflects the required approximation order $k$ and controls how many components are effectively used to describe the data. 

If the variations in the spectral shape due to stellar variability are small, all spectra should be fairly similar. In this case, the most significant difference between measurements would be due to the Doppler shift, which manifests as a complex phase multiplying the principal kernels. In this case, the factorization should result in one singular value significantly larger than all others. 

\subsection{Phase degeneracy}
Presumably, by tracing the entries of $\mathbf{U}_\zeta$ that correspond to $\tilde{U}^{\scaleto{(0)}{5pt}}_t$, the Fourier transform of the zeroth-order principal kernel can be retrieved. While SVD provides a clear framework for decomposing spectra, it presents several challenges if applied na\"ively. 

One such challenge is the inherent ambiguity in the phases of the decomposed singular vectors. While the singular values are uniquely determined, their associated singular vectors are not. This is because the left and right singular vectors can be multiplied by an arbitrary complex phase of the form $e^{i\theta}$ without changing their product. This implies that the left singular vectors, obtained via SVD per-$\zeta$, contain principal kernels multiplied by an array of arbitrary complex phases. 
The Doppler shift in the Fourier domain also appears as a complex phase that depends on the variable $\zeta$. The phase ambiguity, therefore, makes it more difficult to extract the radial velocity. 

This limitation offers a perspective on the problem definition: inferring radial velocity is equivalent to a phase retrieval problem \citep[e.g.,][]{Jaganathan2015}. The phase degeneracy also implies another, more fundamental limitation. Since the Doppler shift is inherently complex, the solution is not unique even if random phases can be accounted for. It can be obtained only up to a multiplicative arbitrary phase. As a result, without a spectral reference, the derived velocities are a relative rather than an absolute measurement. Similar to what happens when extracting the radial velocity using one of the measurements as a template.

\begin{figure*}
        \centering         \includegraphics[width=0.95\textwidth, trim={0cm 0cm 0cm 0cm},clip]{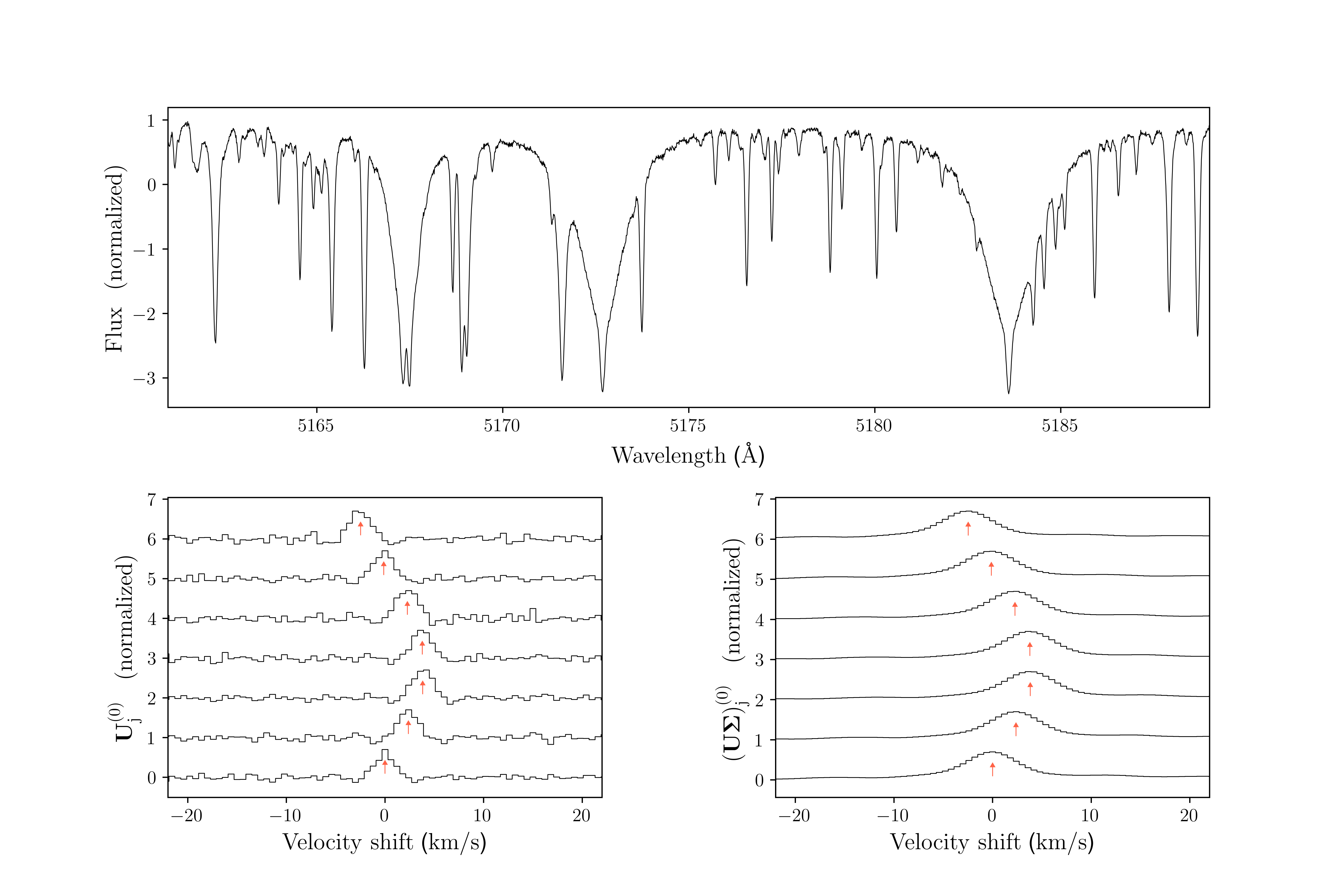} %
        \caption{\textit{Top panel}---a selected segment from one of the simulated spectra used in Section~\ref{sec: phase ratios}, centered around the MgB lines ($T_{\rm eff}=5{,}800 \, {\rm K}$; $\log g = 4.5$; ${\rm[Fe/H]}=0$; $v \sin i = 2\,\,{\rm km\, s}^{-1}$; and $\mathcal{R}=10^5$). \textit{Bottom left panel}---the zeroth-order estimated of the principal kernels obtained from $\mathbf{U}_\zeta$, for the first seven spectra. The kernels are normalized such that their peak value is one and sorted one on top of the other. The arrows represent the Doppler shift used for each spectrum.
        \textit{Bottom right panel}---same as the bottom-left panel, but using the reweighted matrices $\mathbf{U}_\zeta \Sigma_\zeta$.}
    \label{fig: U0 SVD}
\end{figure*}

\section{Radial velocity extraction}
\label{sec: phase retrieval}
This section presents the radial velocity inference scheme. We start by demonstrating the decomposition to principal kernels and formalizing the radial velocity extraction as the complex argument of the ratio between two left singular vectors. We then discuss the radial velocity extraction procedure in an idealized white-noise limit. Finally, we extend the discussion to consider stellar activity.

\subsection{Decomposed Principal Kernels}
As a first step, we demonstrate the decomposition using a spectroscopic time-series based on a synthetic PHOENIX \citep{husser13} template of a Sun-like star. The models were log-uniformly sampled between 5100 and 6100 \r{A}, with spacing equivalent to $600$ m\,s\textsuperscript{-1}.
The instrumental line spread function was assumed to be Gaussian, corresponding to a spectral resolution of $\mathcal{R}=10^5$. A total of 100 synthetic spectra were generated, where each spectrum was assigned a radial velocity according to 
\begin{equation*}
v_\tau = 4 \times \sin \bigg(\frac{\pi}{5} \tau\bigg) \,\,\,{\rm km\, s^{-1}},
\end{equation*}
where $\tau$ is a running index of the simulation. We added Gaussian white noise to the shifted spectrum, with a signal-to-noise ratio of 30 (SNR henceforth; calculated as defined in \citetalias{Shahaf2023}). An example spectrum is presented in the top panel of Figure~\ref{fig: U0 SVD}.

We derive the zeroth-order principal kernels of the simulated spectroscopic time series. To do so, STFT is applied by dividing each spectrum into partially overlapping segments of 4096 bins, with a Hann window applied. The overlap window is selected to ensure that STFT is a tight frame (see the discussion in \citetalias{Shahaf2023}). The matrix $\mathbf{U}_\zeta$ was obtained for each inverse-wavelength using SVD, as equation~(\ref{eq: S SVD}) prescribes. The size of each matrix is $100\,\times\,100$, and its first column provides an estimate of $\tilde{U}^{{\scaleto{(0)}{5pt}}}_t(\zeta)$. 

We apply the inverse Fourier transform on the derived kernels for illustrative purposes. The first seven kernels are shown in the bottom panel of Figure~\ref{fig: U0 SVD}, with an arrow indicating the simulated Doppler shift, where their peak value is expected to appear. The figure suggests that the SNR of the reconstructed kernels is significantly lower than that of the input spectrum. This is largely because unitarity is enforced on the singular vectors and decomposition is done for each $\zeta$ independently. To visually rectify this, one could use $\mathbf{U}_\zeta \Sigma_\zeta$, as the bottom-right panel of Figure~\ref{fig: U0 SVD} illustrates.

\subsection{Phase ratios}
\label{sec: phase ratios}
If the noise of the spectroscopic time series is white and Gaussian, we expect to have only one significant principal spectrum. To illustrate the relative contribution of each order, we plot in Figure~\ref{fig: singular values vs zeta} the first three singular values as a function of $\zeta$ where, for reasons discussed below, we show the squared values of singular values multiplied by the inverse wavelength (namely, $\zeta^2\Sigma^2$). The figure shows that the decomposition is dominated by one component, as expected. The power in the zeroth-order spectrum declines towards high frequencies and drops below the noise level at $\zeta\,{\sim}\,0.25$ s\,km\textsuperscript{-1}. This drop reflects the broadening of the spectrum, due to the assumed rotation and line-spread-function, effectively acting as a low-pass filter.

Equation~(\ref{eq: S bilinear}) illustrates that principal kernels of the same order share the random phases applied to each inverse wavelength. Therefore, random phases should not contaminate the ratio of the two derived principal kernels. This is demonstrated in the bottom panel of Figure~\ref{fig: singular values vs zeta}, showing that the phase information is related to the relative radial velocity. The figure suggests that the relative velocity between each pair of observations can be derived using the phase ratios. However, while the figure illustrates that the information exists, it does not provide an immediate path to obtaining it. We do so by considering the ratios between principal kernels. The relation between the argument of this ratio and the velocity difference between the two observations is 
\begin{equation}
    2\pi \zeta v = - \arg\frac{\tilde{U}^{{\scaleto{(0)}{5pt}}}_{t_i}}{\tilde{U}^{{\scaleto{(0)}{5pt}}}_{t_j}} \simeq - \arg\frac{\mathbf{U}_\zeta[{i,0}]}{\mathbf{U}_\zeta[{j,0}]}\,.
\label{eq: U0 phase ratio}
\end{equation}
The term in the middle represents the theoretical kernel, as shown in equation~(\ref{eq: S t Lambda}), and the term on the right shows the corresponding entries of the SVD matrix obtained. For discussion regarding the impact of measurement noise on the difference between the expected and measured kernels, see Section~\ref{sec: kernel leakage}.

The top panel of Figure~\ref{fig: singular values vs zeta} shows some excess in the first-order kernel. In this case, the excess is due to cross-talk between the Doppler shift and STFT window function. The simulation shown in Figures~\ref{fig: U0 SVD} and~\ref{fig: singular values vs zeta} assumes a radial velocity semi-amplitude of a few kilometers per second. This is a large modulation compared to the width of the segment size and STFT window function. This effect vanished for smaller radial velocity semi-amplitudes or larger segment sizes, as this error scales with the ratio between the radial velocity and the width of the segment. This plot was generated with segment sizes large enough so the effect is visible, yet mild. We emphasize that this work is focused on sub-meter per second velocity modulations, in which case the impact of this artifact is negligible.

\subsection{White noise limit}
The ratio between the kernels reflects the radial velocity difference, as equation~(\ref{eq: U0 phase ratio}) shows. At each inverse wavelength, $\zeta$, the kernel values are assumed to follow a complex normal distribution. While the ratio of two complex Gaussian variables is generally non-Gaussian, in the high signal-to-noise ratio (SNR) limit the problem approaches Gaussian statistics. See, for example, \citet[][]{Sourisseau2019}. 

\subsubsection{Velocity extraction}
We start by considering the velocity difference between two epochs at a given inverse wavelength, $\zeta$. We treat the first measurement as a reference and estimate the radial velocity of the second measurement using the Fourier transform of their corresponding zeroth-order principal kernels. The reference and comparison kernels are subject to measurement errors, expressed as
\begin{equation}
\begin{aligned}
    {\mathbf{U}}_{_\zeta}^{\tiny \rm ref} \,\,\,&= z + \varepsilon_{\tiny \rm r}, \quad \text{and}\\[4pt]
    {\mathbf{U}}_{_\zeta}^{\tiny \rm cmp} &= z \exp\big({-i2\pi\zeta v}\big) + \varepsilon_{\tiny \rm c},
\end{aligned}
\end{equation}
where $z$ is the nominal value of the reference kernel at $\zeta$, $v$ is the velocity difference between the two measurements, and $\varepsilon_{\tiny \rm r}$ and $\varepsilon_{\tiny \rm c}$ are the measurement errors associated with the reference and comparison kernels, respectively.
For brevity, we take the reference as the first zeroth-order kernel and the comparison term as the zeroth-order kernel of the $i$\textsuperscript{th} measurement. 
If $\max(|\varepsilon_{\tiny \rm r}|,|\varepsilon_{\tiny \rm c}|) \ll |z|$, the kernel ratio can be approximated to first order, 
\begin{equation}\label{eq: U0 ratio}
    \mathcal{R}^{(0)}_{_{\zeta i}} = \frac{{\mathbf{U}}_{_\zeta}[i,0]}{{\mathbf{U}}_{_\zeta}[0,0]} \simeq \exp\big({-i2\pi\zeta v_i}\big) + \varepsilon_{_\zeta},
\end{equation}
where $v_i$ is the radial velocity difference between the $i$\textsuperscript{th} and $0$\textsuperscript{th} epochs, and we defined $\varepsilon_{_\zeta} \equiv (\varepsilon_{\tiny \rm c} - \varepsilon_{\tiny \rm r})\,z^{-1}$. 

We assume that the errors are independent circularly symmetric complex Gaussian random variables. In the high-SNR limit considered here, $\varepsilon_{_\zeta}$ is also circularly symmetric, with variance inversely proportional to the SNR squared. The ratio distribution, therefore, attains the form
\begin{equation}\label{eq: R complex normal}
    \mathcal{R}^{(0)}_{_{\zeta i}} \sim \mathcal{CN}\big\{\exp({-i2\pi\zeta v_i}); \,\, 2\,{\sigma}_{_\zeta}^{2}\big\},
\end{equation}
where ${\sigma}_{_\zeta}^{-1}$ is the SNR at the specific $\zeta$ and we assumed that the SNRs of both kernels are similar.
The maximum likelihood estimator of the velocity term in equation~(\ref{eq: R complex normal}) is given by
\begin{equation}\label{eq: v zeta}
    \hat{v}_{_{\zeta i}} = -\frac{1}{2\pi\zeta}\arctan\left[\frac{\Im({\mathcal{R}^{(0)}_{_{\zeta i}}})}{\Re({\mathcal{R}^{(0)}_{_{\zeta i}}})}\right] \simeq -\frac{1}{2\pi\zeta}\frac{\Im({\mathcal{R}^{(0)}_{_{\zeta i}}})}{\Re({\mathcal{R}^{(0)}_{_{\zeta i}}})},
\end{equation}
where in the last transition, we assumed that $\zeta v_i$ is close to zero.
The Fisher information is
\begin{equation}\label{eq: I zeta}
    \mathcal{I}_{_{\zeta i}} \propto \frac{\zeta^2}{\sigma_{_\zeta}^2} \mathbb{E}\left[{\Re\left({\mathcal{R}^{(0)}_{_{\zeta i}}} e^{-i2\pi\zeta v_i}\right)} \right] \rightarrow \frac{\zeta^2}{\sigma_{_\zeta}^2}.
\end{equation}
Since $\sigma_{_\zeta}$ is the SNR at a given $\zeta$, this equation implies that Fisher information scales with $(\zeta \times \text{SNR})^2$. This reflects a balance between the radial velocity signal, which has a more substantial effect at larger inverse wavelengths, and the instrument's finite resolution, which acts as a low-pass filter. Therefore, even if the SNR may be high at small $\zeta$, the significant of the measurement in this regime is small. Conversely, at sufficiently large $\zeta$ the signal is depleted, and the measurement is dominated by noise.

\begin{figure}
        \centering 
        \includegraphics[width=0.45\textwidth]{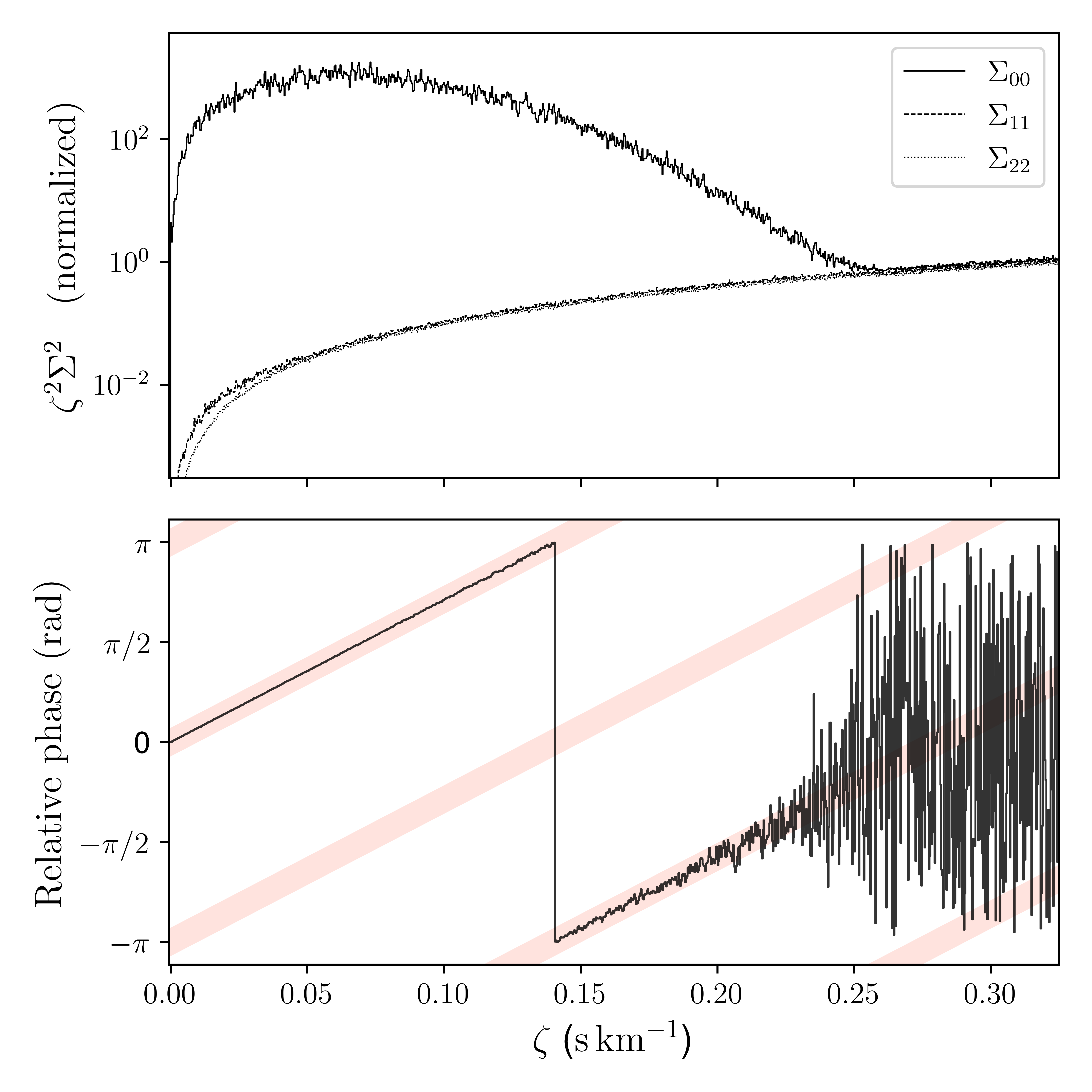} %
        \caption{\textit{Top panel}---the normalized information in the first three principal spectra (corresponding to $\Sigma_{00}$, $\Sigma_{11}$ and $\Sigma_{22}$) versus their corresponding inverse-wavelength, $\zeta$, for the simulated white-noise spectroscopic time-series. All values are normalized to $\Sigma_{00}$ at $\zeta=0$.  \textit{Bottom panel}---relative phases between the first and fourth simulated spectra, shown as a solid black line. The red stripes correspond to the expected relative phase due to the velocity difference between the two observations, i.e., $2\pi\,\zeta\,\times\,(v_4-v_1)$. Notably, the relative phase follows the expected trend for inverse wavelength below $\sim 0.2$ s\,km\textsuperscript{-1}, where the information content falls below the noise level.}
    \label{fig: singular values vs zeta}
\end{figure}
The velocity of the $i$\textsuperscript{th} measurement at a given $\zeta$ is given in equation~(\ref{eq: v zeta}). Since the kernels for all measurement epochs are sampled on the same $\zeta$-grid, the velocity constraints from different inverse wavelengths can, in principle, be combined into a single system of equations. One can therefore solve for all velocities simultaneously by assembling the per-$\zeta$ constraints into a matrix equation and solving the corresponding normal equations, as shown in Appendix~\ref{app: matrix form}. In practice, this direct approach can be challenging, because the resulting matrices are large and sparse, and numerical stability may become a limiting factor at the precision required here.

As a simple alternative, one can use the inverse-variance weighted velocity, obtained from of all estimates of $\hat{v}_{_{\zeta i}}$. For this purpose, we estimate the velocity uncertainty using equation~(\ref{eq: I zeta}) and the Cram\'er-Rao lower bound. The inverse-variance weight of $v_{_{\zeta i}}$ is given by Fisher information of the corresponding $\zeta$, divided by the total information of all inverse wavelengths, namely
$\mathcal{I}_{_\zeta}/\sum{\mathcal{I}_{_\zeta}}\equiv \mathcal{I}_{_\zeta}/\mathcal{I}_{\rm tot}$. As a result, the velocity estimate of the i\textsuperscript{th} epoch becomes
\begin{equation}\label{eq: v hat}
    \hat{v}_i \simeq \sum_\zeta  \frac{\mathcal{I}_{_\zeta}}{\mathcal{I}_{\rm tot} } \hat{v}_{_{\zeta i}}\, .
\end{equation}

\begin{figure}
        \centering \includegraphics[width=0.45\textwidth]{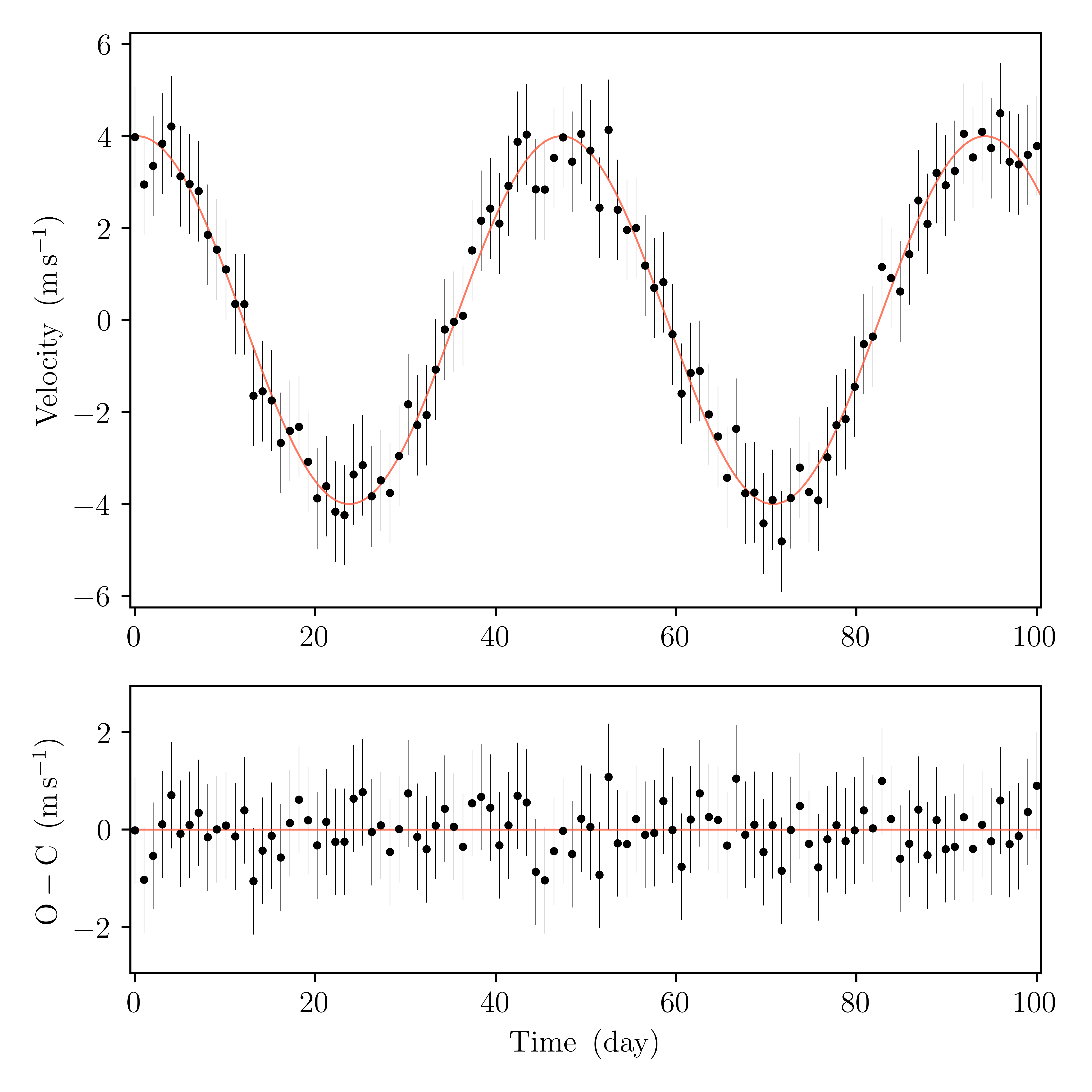} %
        \caption{\textit{Top panel}---Velocities extracted for the white noise simulation. The simulated spectroscopic dataset is based on a synthetic spectrum of a Sun-like star, and the noise is assumed to be white and Gaussian, with an SNR of 30. The injected radial velocity signal is a solid red curve in the background.  \textit{Bottom panel}---The residuals between the injected and inferred velocities. The scatter of the residuals is 58~cm~s\textsuperscript{-1}, consistent with the estimated error.}
    \label{fig: RV extraction white noise}
\end{figure}

\subsubsection{Noise estimates}

Equation~(\ref{eq: I zeta}) shows that the information scales inversely with $\sigma_{_\zeta}^2$, where $\sigma_{_\zeta}$ is the standard deviation of the complex kernel ratio at inverse wavelength $\zeta$. Since $\sigma_{_\zeta}^{-1}$ is equivalent to the SNR,  we can  estimate the information we need to quantify the SNR per $\zeta$. We do so using the singular values themselves, via
\begin{equation}\label{eq: snr estimate}
{\rm SNR} \simeq \sigma_\zeta^{-1} \propto \frac{\Sigma_{00}}{\Sigma_{\infty}},
\end{equation}
where $\Sigma_{00}$ is the leading singular value and $\Sigma_{\infty}$ is the leading singular value expected from a purely random matrix, proportional to the noise floor level. It is estimated from the cutoff beyond which the signal becomes subdominant and $\Sigma_{00}$ drops below $\Sigma_{\infty}$. For example, in Figure~\ref{fig: singular values vs zeta}, we estimate $\Sigma_{\infty}$ as the singular value at $\zeta \sim 0.25$~s~km$^{-1}$. 

Using the SNR scaling relation above, we derive the Fisher information at each inverse wavelength using equation~(\ref{eq: I zeta}),
\begin{equation}\label{eq: info estimate}
\mathcal{I}_{_\zeta} \propto \zeta^2 \left(\frac{\Sigma_{00}}{\Sigma_{\infty}}\right)^2.
\end{equation}
The radial velocity uncertainty is inversely proportional to the square root of the total information, so that 
\begin{equation}\label{eq: delta v white noise}
\Delta \hat{v}_i \propto \Sigma_{\infty} \left( \sum_\zeta \zeta^2 \Sigma_{00}^2 \right)^{-1/2}.
\end{equation}
This error estimate is given up to the scaling coefficient of $\Sigma_\infty$, and factors omitted for simplicity. 

Using the asymptotic expression from \citet{Vershynin2018}, we obtain $\Sigma_\infty / \sigma_\star \approx  \left(\sqrt{n} + \sqrt{m}\right)$, where $\sigma_\star$ is the Gaussian noise level, $n$ is the number of observations, and $m$ is the number of spectral segments. However, the zeroth-order singular value is also likely to increase with the number of measurements. This is because the signal is expected to behave coherently between matrix entries, especially for the zeroth-order-term. Therefore, we assume that $\Sigma_{00}\propto\sqrt{n}$.
Making the required replacements, rewrite the error estimate as  
\begin{equation}\label{eq: delta v white noise - verishnin}
\Delta\hat{v}_i \sim  \frac{\Sigma_\infty}{\pi} \frac{\sqrt{n} }{\sqrt{n} + \sqrt{m}}\left(\sum_\zeta \zeta^2 \Sigma_{00}^2\right)^{-1/2}.
\end{equation}
We note that the equation above should be treated as a rough estimate of the noise budget, as it relies on our estimate of $\Sigma_\infty$ and its asymptotic relations to the noise. Improved estimates of the the noise level and, therefore, the radial velocity error can be rescaled post-priory if needed (see below).

\subsubsection{A Simple Example}
Figure~\ref{fig: RV extraction white noise} illustrates the velocity extraction procedure in the presence of white Gaussian noise. The time series is generated using the same synthetic model parameters described in Section~\ref{sec: phase retrieval}, except for the radial velocity semi-amplitude. To ensure the validity of the linear approximation, the velocity shift is limited to a few meters per second. We generate 100 synthetic spectra, assuming the sampling is evenly spaced over a 100-day baseline. The injected period is of 47 day and, compared to the previous examples, we reduced the radial velocity semi-amplitude by three orders of magnitude to $ 4~\,{\rm m\, s^{-1}}$. The segment length was taken to be $2^{10}$ bins with 75 percent overlap. This selection yields a number of independent segments close to the number of measurements, which we found to be a favorable selection.

The measurement error per $\zeta$ is inversely proportional to the SNR, as equation~(\ref{eq: R complex normal}) shows. Here ${\Sigma_{\infty}}$ represents the asymptotic singular values due to the noise level. Here, we approximate $\Sigma_\infty$ using the zeroth-order singular value, $\Sigma_{00}$, at an inverse wavelength of $0.25$~s~km\textsuperscript{-1}, and calculate the Fisher information according to equation~(\ref{eq: info estimate}). As discussed above, this is the point where the principal values converge to the baseline noise level (see Figure~\ref{fig: singular values vs zeta}). To ensure the high-SNR approximation in equation~(\ref{eq: R complex normal}) holds, only $\zeta$'s with SNR larger than 10 are included in the analysis. 

The inferred velocities and residuals are shown in Figure~\ref{fig: RV extraction white noise}. The errors appear slightly overestimated, which follows from the $\zeta$ uncertainty scaling in equation~(\ref{eq: snr estimate}). Factors of order unity in that relation propagate directly into the velocity uncertainties. This limitation, inherent to the approximate error estimate used here, affects only the error bars. The velocities themselves remain unbiased.

\subsection{Activity-induced signal}
\label{sec: kernel leakage}
Ideally, variability that is not related to the Doppler shift would be captured entirely by higher-order kernels, leaving the zeroth-order sensitive only to velocity. In practice, any process that modifies line shapes across epochs introduces some overlap with the Doppler component. 
Examples include magnetic activity, convection, oscillations, micro-telluric absorption, reduction systematics, and wavelength-calibration drifts. Because the decomposition is performed on a finite and noisy matrix, this variability is not confined to the higher-order kernels and a small fraction leaks into the lower-order ones. This cross-talk introduces a corresponding level of contamination in the inferred velocities. The remainder of this section quantifies this effect and shows how it can be corrected.

\subsubsection{Kernel cross-talk}

To leading order, we assume that each principal kernel is contaminated by its closest neighbor of higher order (see a justification Appendix~\ref{app: perturbed SVD}). We express the relation between the singular vectors obtained through SVD and our expected principal kernels as
\begin{equation}\label{eq: U0 perturbed}
{\mathbf{U}}_\zeta[i,0] \simeq  \tilde{U}^{{\scaleto{(0)}{5pt}}}_i + \epsilon\, \tilde{U}^{{\scaleto{(1)}{5pt}}}_i,
\end{equation}
where $\epsilon$ represents the perturbation coefficient. In the following, the magnitude and phase of the perturbation coefficients are taken somewhat loosely. It only represents some small number, close to zero, that can generally be a function of $\zeta$. Although it is generally complex, we note that we do not know that phase of the principal kernels, so we assume for simplicity that $\epsilon$ is real and absorb its complex phase into the kernel.

Equation~(\ref{eq: U0 perturbed}) suggests that the observed ratio between the comparison and reference kernels is given by 
\begin{equation} \label{eq: obs ratios}
    \begin{aligned}
    \hat{\mathcal{R}}^{(0)}_{_{\zeta i}}  & \equiv \frac{{\mathbf{U}}_\zeta[i,0]}{{\mathbf{U}}_\zeta[0,0]} =\frac{ \tilde{U}^{{\scaleto{(0)}{5pt}}}_i}{ \tilde{U}^{{\scaleto{(0)}{5pt}}}_0 + \epsilon\,\tilde{U}^{{\scaleto{(1)}{5pt}}}_0} +  \epsilon\, \frac{   \tilde{U}^{{\scaleto{(1)}{5pt}}}_i}{ \tilde{U}^{{\scaleto{(0)}{5pt}}}_0 + \epsilon\,  \tilde{U}^{{\scaleto{(1)}{5pt}}}_0}\\[5pt]
    &\simeq \mathcal{R}^{(0)}_{_{\zeta i}}  \left[1+\epsilon\,\frac{
\tilde{U}^{{\scaleto{(1)}{5pt}}}_0}{\tilde{U}^{{\scaleto{(0)}{5pt}}}_0}\right]^{-1} + \mathcal{R}^{(1)}_{_{\zeta i}} \left[1+\epsilon^{-1}\,\frac{
\tilde{U}^{{\scaleto{(0)}{5pt}}}_0}{\tilde{U}^{{\scaleto{(1)}{5pt}}}_0}\right]^{-1} \\[5pt]
    &\simeq \mathcal{R}^{(0)}_{_{\zeta i}}  \big(1-\epsilon\big) + \epsilon\, \,\mathcal{R}^{(1)}_{_{\zeta i}} ,
    \end{aligned}
\end{equation}
where we assumed that ${
|\tilde{U}^{{\scaleto{(1)}{5pt}}}_0}/{\tilde{U}^{{\scaleto{(0)}{5pt}}}_0}|\lesssim1$ and $|\epsilon_0| \simeq |\epsilon_i| \ll 1$, in the range of inverse wavelengths considered. The equation above shows how rotation of the principal kernels propagates to the kernel ratios and, through it, to the inferred velocities. Following the same steps shown in equation~(\ref{eq: U0 phase ratio}), we obtain a relation between the observed velocity and the expected one, 
\begin{equation}
\begin{aligned} \label{eq: v contamination}
    \hat{v}^{\rm obs}_{\zeta} &\simeq -\frac{1}{2\pi \zeta} \frac{ 
    \Im(\hat{\mathcal{R}}^{(0)})}{\Re(\hat{\mathcal{R}}^{(0)})} \\[5pt]
    &\simeq -\frac{1}{2\pi\zeta}\left[(1-\epsilon \big)\frac{ 
    \Im({\mathcal{R}}^{(0)})}{\Re({\mathcal{R}}^{(0)})}  + \epsilon \frac{ 
    \Im({\mathcal{R}}^{(1)})}{\Re({\mathcal{R}}^{(1)})} \right] \\[5pt]
    & \equiv(1-\epsilon) \hat{v}_{_{\zeta i}}^{(0)} + \epsilon \,\hat{v}_{_{\zeta i}}^{(1)}.
\end{aligned}
\end{equation}
Assuming that the perturbation on the first-order kernel ratio is small, the parasitic signal $\hat{v}_{_{\zeta i}}^{(1)}$ can be estimated directly from ${{\mathbf{U}}_\zeta[i,1]}/{{\mathbf{U}}_\zeta[0,1]}$, as systematic deviations of it from the ideal unperturbed kernel ratio will appear as a second-order effect. An application of correction is demonstrated in the following.

To scale the impact of the kernel cross-talk on the radial velocity precision, we treat the $\epsilon \,\hat{v}_{_{\zeta i}}^{(1)}$ term as contamination-induced noise. The quadrature sum of the error contributed by the contaminants in all inverse wavelengths is proportional to $\sqrt{\sum \zeta^{-2}\epsilon^2}$. Plugging in the expected scaling relation for the leakage, shown in equation (\ref{eq: leak bound}), we estimate the contamination from the high order terms as
\begin{equation}
    \Delta \hat{v}_i^{\rm c} \propto \frac{1}{\sqrt{n}}\left[\sum_\zeta  \frac{1}{\zeta^2}\left(\frac{\Sigma_{11}}{\Sigma_{00}}\right)^2\right]^{0.5} \lesssim \frac{\max{\Sigma_{11}}}{\sqrt{n}} \left(\sum_\zeta \zeta^2 \Sigma_{00}^2\right)^{-1/2}.
\end{equation}
The ratio between this activity-induced variance and the white noise scaling from equation~(\ref{eq: delta v white noise}) yields
\begin{equation}\label{eq: noise contamination error}
    \Delta \hat{v}_i^{\rm c} /\Delta \hat{v} \lesssim
    \frac{\pi}{2}\,\sqrt{\frac{1}{n}} \,
    \frac{\max{\Sigma_{11}}}{\Sigma_\infty},
\end{equation}
where we assumed that $m\simeq n$.

\begin{figure}
        \centering \includegraphics[width=0.85\columnwidth]{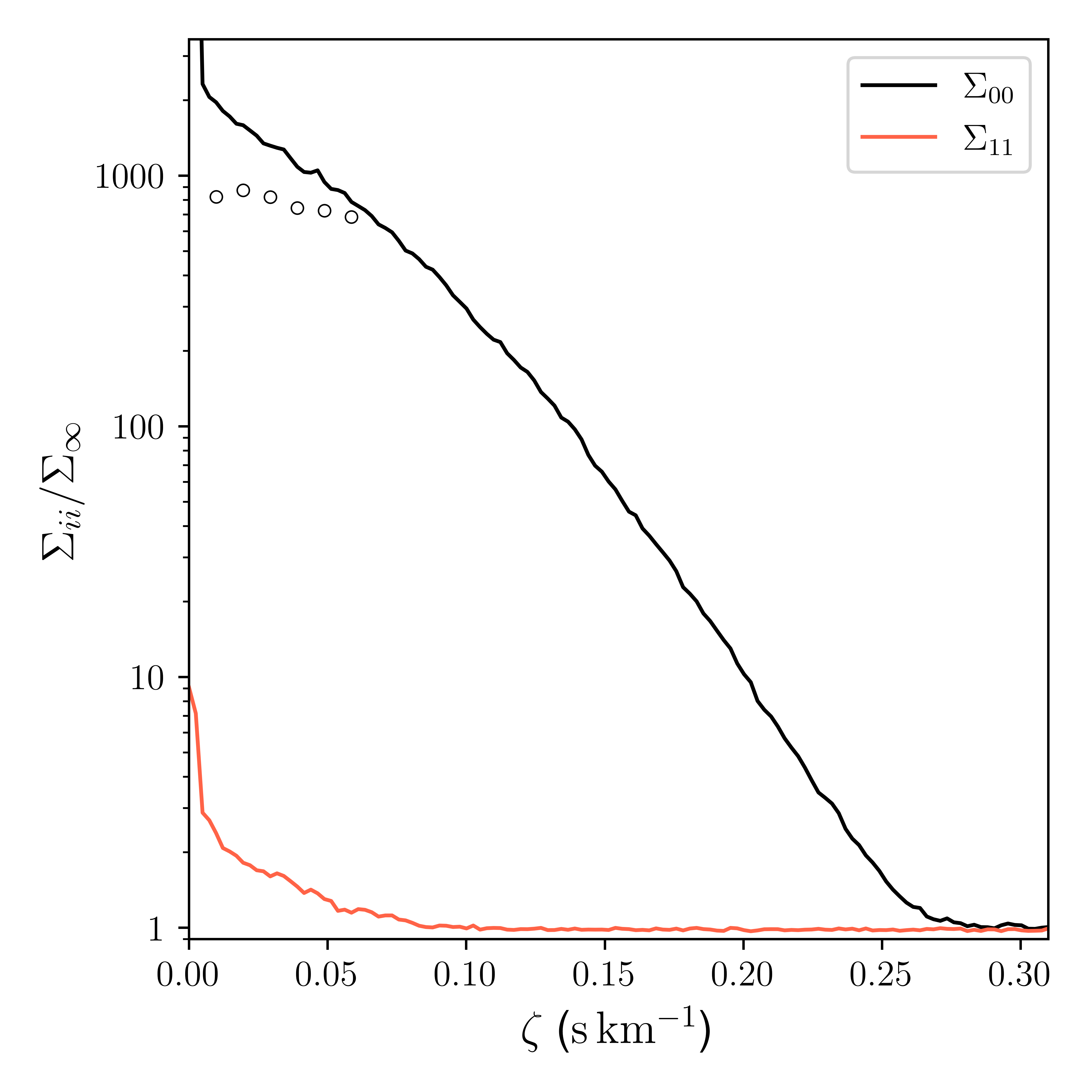} %
        \caption{The normalized zeroth- and first-order principal values versus inverse wavelength, for the SOAP simulation presented in Section~\ref{sec: SOAP}. The normalization constant given by an estimate of $\Sigma_\infty$, estimated as the median singular value for $\zeta > 0.3$ s~km\textsuperscript{-1}. The white dots show the ratio between the zeroth and first order terms, i.e., $\Sigma_{00}/\Sigma_{11}$.}
    \label{fig: principal values soap}
\end{figure}

\begin{figure*}
        \centering 
        \includegraphics[width=0.975\textwidth]{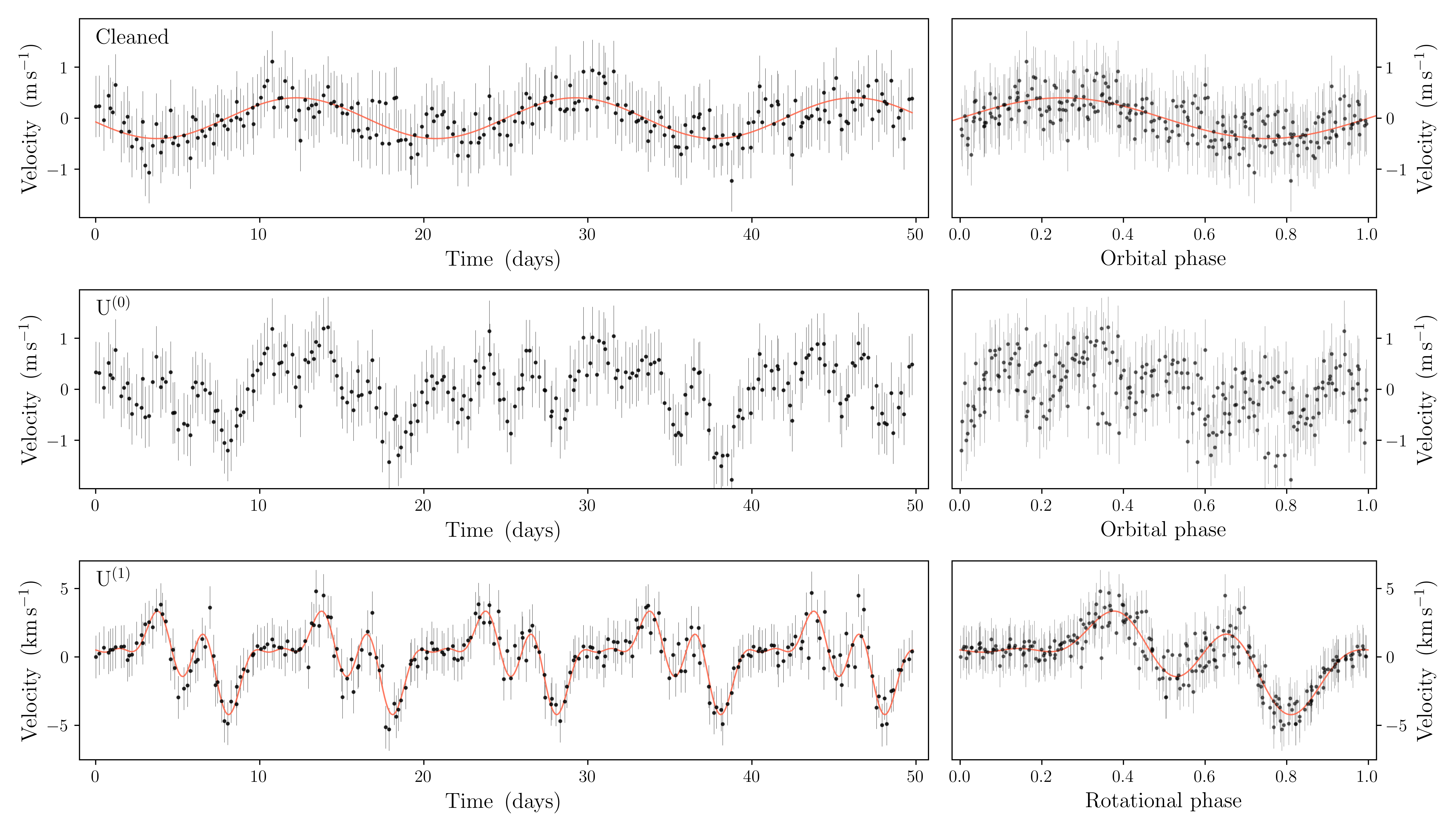} %
        \caption{\textit{Top panels}---The rectified velocities, $v_{\rm c}$, extracted for the SOAP simulation described in Section~\ref{sec: SOAP}. The simulated spectroscopic dataset is based on a synthetic spectrum of a Sun-like star, and the noise is assumed to be white and Gaussian. The injected radial velocity signal is a solid red curve.  
        \textit{Middle panels}---the extracted velocities from the zeroth-order kernel, $U^{(0)}$. These velocities show a combination of the injected Doppler shift and parasitic signal induced by stellar activity.
        \textit{Bottom panels}---the extracted velocities from the first-order kernel, $U^{(1)}$, dominated by stellar activity. The solid red curve represents the fitted 5-harmonic model fitted to the extracted velocities (see text).}
    \label{fig: RV extraction soap}
\end{figure*}

\subsubsection{SOAP simulation}
\label{sec: SOAP}
  We use the same SOAP-GPU \citep{zhao23, Zhao2025} simulation from \citetalias{Shahaf2023} to demonstrate the performance of the SVD-based extraction method and discuss this issue. 

The simulation is based on a quiet Sun spectrum, with spots and faculae modeled using PHOENIX templates at $5015$ and $6028$ K, respectively. Spot sizes and faculae-to-spot size ratios were randomly drawn from \citet[][Table 1]{borgniet15}. The latitude distribution follows the butterfly diagram, with longitudes separated by $180^\circ$, reproducing the observed active longitudes. Active regions were incorporated while accounting for convective blueshift inhibition and the dependence of the line bisector on the center-to-limb angle, $\mu$, with no size evolution applied.

The synthetic dataset consists of 249 spectra spanning 50 days, with an average cadence of 5 hours. The simulated star has a 10-day rotation period and an average of 10 active regions, with a median spot size of ${\sim} 50$-millionth of a solar hemisphere and an interquartile range of ${\sim}15{-}250$ millionths. The expected active region filling factor is ${\sim}1\%$. To simulate planetary reflex motion, we introduced a periodic velocity shift of
\begin{equation*}
    v = 30 \times \sin\bigg[ \frac{2\pi\, (t-t_0)}{17~{\rm day}} \bigg] \,\,\text{cm\,s\textsuperscript{-1}},
\end{equation*}
where $t_0$ was taken as the median of the simulated temporal baseline. The spectrum was sampled on a logarithmic grid with a 800 m~s\textsuperscript{-1} step size over a relatively broad bandpass of $5000-6500$ angstroms. Gaussian white noise was added to the spectrum, assuming an SNR of 65 which, by the definitions in \citetalias{Shahaf2023}, is roughly equivalent to SNR 250 from photon counts at 550 nm. 

We extracted the velocities from the simulated dataset, using segment length of $2^{9}$ bins and overlap of 75 per-cent, aiming to set the number of independent segments close to the number of observations. Figure~\ref{fig: principal values soap} shows that the zeroth order principal values reach the noise floor at  $\zeta\,{\gtrsim}\,0.30$~s~km\textsuperscript{-1}, therefore, we estimate $\Sigma_\infty$ as the median value of $\Sigma_{00}$ in this range. The impact of stellar activity can be seen from the shape of $\Sigma_{11}$, plotted as a red curve in the same figure. The perturbation is small, and drops below the noise level at $\zeta\sim0.1$~s~km\textsuperscript{-1}. Using equation~(\ref{eq: noise contamination error}), we estimate that the leak is comparable to the shot-noise level, i.e., $\Delta \hat{v}_i^{\rm c}\approx \Delta \hat{v}$. 

The leak from the first order kernel should be fitted and accounted for. Therefore, unlike the purely Gaussian white noise case, we extract velocities twice: first using zeroth-order kernels and then first-order ones. These velocities are presented in the middle and bottom panels of Figure~\ref{fig: RV extraction soap}. For the zeroth order velocity, $\hat{v}^{(0)}$, we use inverse wavelengths such that $\Sigma_{00}/\Sigma_\infty > 10$ and extract the velocities as described in equations~(\ref{eq: v hat}) and~(\ref{eq: info estimate}). For the first-order velocities, $\hat{v}^{(1)}$, we use all entries such that $\zeta<0.1$~s~km\textsuperscript{-1}, since $\Sigma_{11}$ has fewer entries above the noise level (see Figure~\ref{fig: principal values soap}). 
Assuming that at $\zeta \simeq 0.1$~s~km\textsuperscript{-1} the contribution of the signal is roughly four times smaller than the white noise level, we estimate $\Sigma_{11}/\Sigma_\infty\propto4\,\sigma_{_\zeta}^{-1}$, and make the adjustment to equation~(\ref{eq: info estimate}) accordingly.

The middle panel of Figure~\ref{fig: RV extraction soap} shows that the zeroth-order velocities are contaminated by parasitic signal due to stellar activity. The bottom panel shows the first-order velocities, which are modulated with the 10-day rotation period of the simulated star with peak-to-peak amplitude of ${\sim}\,5$~km\,s\textsuperscript{-1}. This modulation leaked to the zeroth-order velocities. To remove this contamination, we fit a 5-harmonic model to $\hat{v}^{(1)}$, using the rotation of the star as the fundamental period. The fitted model is plotted in the bottom panels of Figure~\ref{fig: RV extraction soap}. We remove the activity-induced signal from $\hat{v}^{(0)}$ by fitting out the 5-harmonic model from it, following the prescription in equation~(\ref{eq: v contamination}). This simple linear fit resulted in
\begin{equation}
    \hat{v} = \hat{v}^{(0)} - \big(292 \pm 23 \,\, {\rm ppm}\big) \hat{v}_{\rm m}^{(1)} - \big(0.6 \pm 3.1 \,\, {\rm cm~s^{-1}}\big)\,,
\end{equation}
where here $\hat{v}^{(0)}$ represents the velocities obtained from the zeroth-order kernels and $\hat{v}_{\rm m}^{(1)}$ is the 5-harmonic model fitted to $\hat{v}^{(1)}$.
The specific choice of modeling approach for $\hat{v}_{\rm m}^{(1)}$ is not crucial. In this example we use a harmonic fit because the contamination is strictly periodic, but a Gaussian Process or any other flexible model could be used instead. The essential point is that the contamination dominates the higher-order kernel, so its temporal behavior can be learned directly from the first-order velocities. This requires an effective decomposition, achievable with sufficient sampling and SNR.

The cleaned velocity time series is shown in the top panel of Figure~\ref{fig: RV extraction soap}. We used the same synthetic data as in \citetalias{Shahaf2023}, achieving comparable results. Figure~\ref{fig: RV extraction soap} provides a demonstration of kernel cross-talk, as Section~\ref{sec: kernel leakage} outlines.  It also shows the ability of to mitigate its impact linearly by using the information provided by the high-order terms in the decomposition. Furthermore, while in \citetalias{Shahaf2023} we had to use a model for the high-order terms, the velocities here were obtained by using only the information provided in the data.

\section{Test cases} \label{sec: test cases}
To test the method under realistic observational conditions, we 
apply it to a spectroscopic time series of HD 34411 and $\tau$~Ceti. Observations were obtained from 2019 through 2020 with the Extreme Precision Spectrograph (EXPRES; \citealt{Jurgenson2016}, \citealt{Blackman2020}) mounted on the 4.3-meter Lowell Discovery Telescope. EXPRES delivers $R\sim137{,}000$ over the $380 - 680$ nm range and is optimized for sub-meter-per-second radial velocity precision. The spectra were processed with the standard pipeline \citep{Petersburg2020}. 

The radial velocities were analyzed and discussed thoroughly by \citet{zhao22}, and their observations were part of the 100 Earths Survey \citep{Brewer2020}. The integration time was designed to achieve SNR of 250 per pixel at 550~nm. The expected shot-noise–limited precision at this resolution is ${\sim}\,30~{\rm cm\,s}^{-1}$. In practice, the total observed scatter is ${\sim}\,50~{\rm cm\,s}^{-1}$, which includes contributions from instrumental stability, calibration, and stellar variability \citep{Petersburg2020, Brewer2020}. On top of this baseline precision, we expect additional scatter due to the temporally resolved sampling of the p-modes, as discussed for each target below.

\subsection{Test case I: HD 34411}

HD~34411 is a quiet G0V star at a distance of ${\sim}\,12.6$ pc. It has an apparent $V$-band magnitude of ${\sim}\,4.7$, an effective temperature of ${\sim}\,5900$ K, and metallicity of about $0.1$ dex. It is a chromospherically quiet, slow rotating star, with an estimated age of ${\sim}\,4.8$ Gyr \citep{zhao22}. The median exposure time of this target is ${\sim}\,2.5~{\rm min}$, which is equivalent to about a third of the stellar p-mode cycle \citep{Brewer2020}. This fractional coverage of the cycle is expected to manifest as additional scatter to the velocities, of up to $100~{\rm cm\,s}^{-1}$ \citep{Chaplin2019}. Considering the additional terms discussed above, we expect the scatter of up to ${\sim}\,1\,{\rm m\,s}^{-1}$.

The dataset includes 188 measurements of HD~34411, taken over a 14-month baseline. Out of these, we consider a subset of 182 measurements with exposure times below $10$~min, to avoid nights of poor observing conditions. We analyze 40 orders (number $93-133$), covering the 460–660 nm range. Each order is resampled onto a log-uniform grid with a velocity spacing equivalent to 600 m\,s$^{-1}$. Only the central regions of each order are retained, where the blaze remains above 30 percent of its peak value. To validate the results we injected a Doppler shift according to
\begin{equation*}
    v = 50 \times \sin\bigg[ \frac{2\pi\, (t-t_0)}{75~{\rm day}} \bigg] \,\,\text{cm\,s\textsuperscript{-1}}.
\end{equation*}
where $t_0$ is the modified Julian day of 59000. The synthetic signal is injected directly into the telluric-corrected spectra. For each exposure, we shift the observed spectrum according to the assumed Keplerian velocity at that epoch and then rebin it onto the original logarithmic wavelength grid.

\begin{figure}
        \centering         
        \includegraphics[width=0.8\columnwidth]{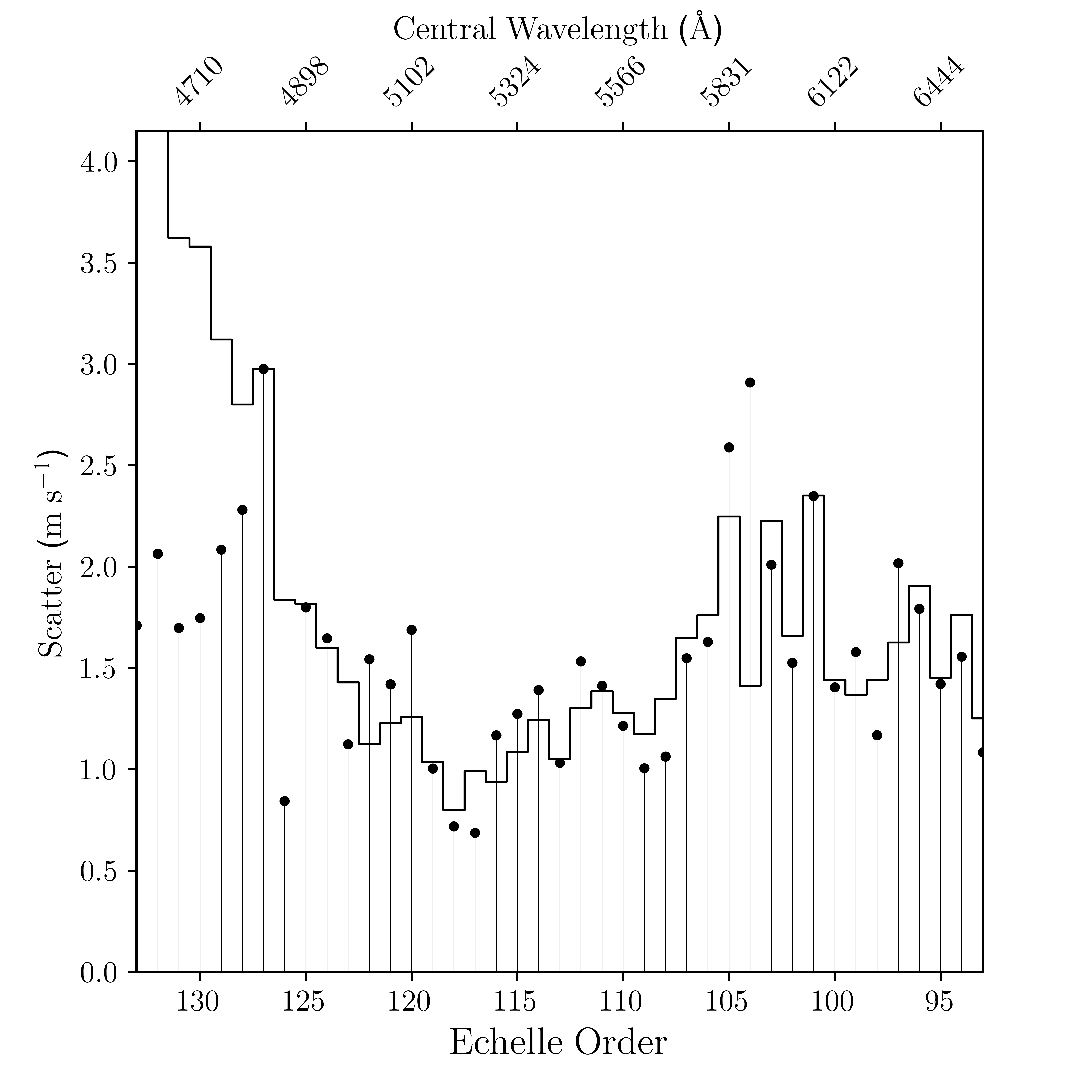} %
        \caption{HD~34411 variance versus order number. The central wavelength of each order is presented at the top panel. The empty step functions shows a Fisher-information-based estimate of the scatter. The stem plot shows the empirical scatter, derived using samples with exposure times of 5-8 min, to mitigate the impact of p-modes (see text). The figure shows that most of the information is due to a few central orders around 5200 Angstroms, with degrading accuracy towards the blue and red parts of the spectrum.}
    \label{fig: HD 34411 order var}
\end{figure}

Orders are treated independently to account for possible variations in the line spread function, reduction systematics, and varying SNR. Since a typical number of wavelength bins in the spectrum is ${\sim}\,5500$, we use wavelength segments of $2^5$ bins, equivalent to $19.2$ km~s\textsuperscript{-1} segment width, with 75\% segment overlap (see \citetalias{Shahaf2023}). This choice ensures that the number of independent STFT segments is close to the number of observations. This is not required by the framework, but we have found that it improves the conditioning of the SVD, since matrices with similar dimensions tend to have more stable singular-value spectra. In practice, this yields a cleaner separation between the dominant and higher-order components.

\begin{figure}
        \centering         
        \includegraphics[width=0.8\columnwidth]{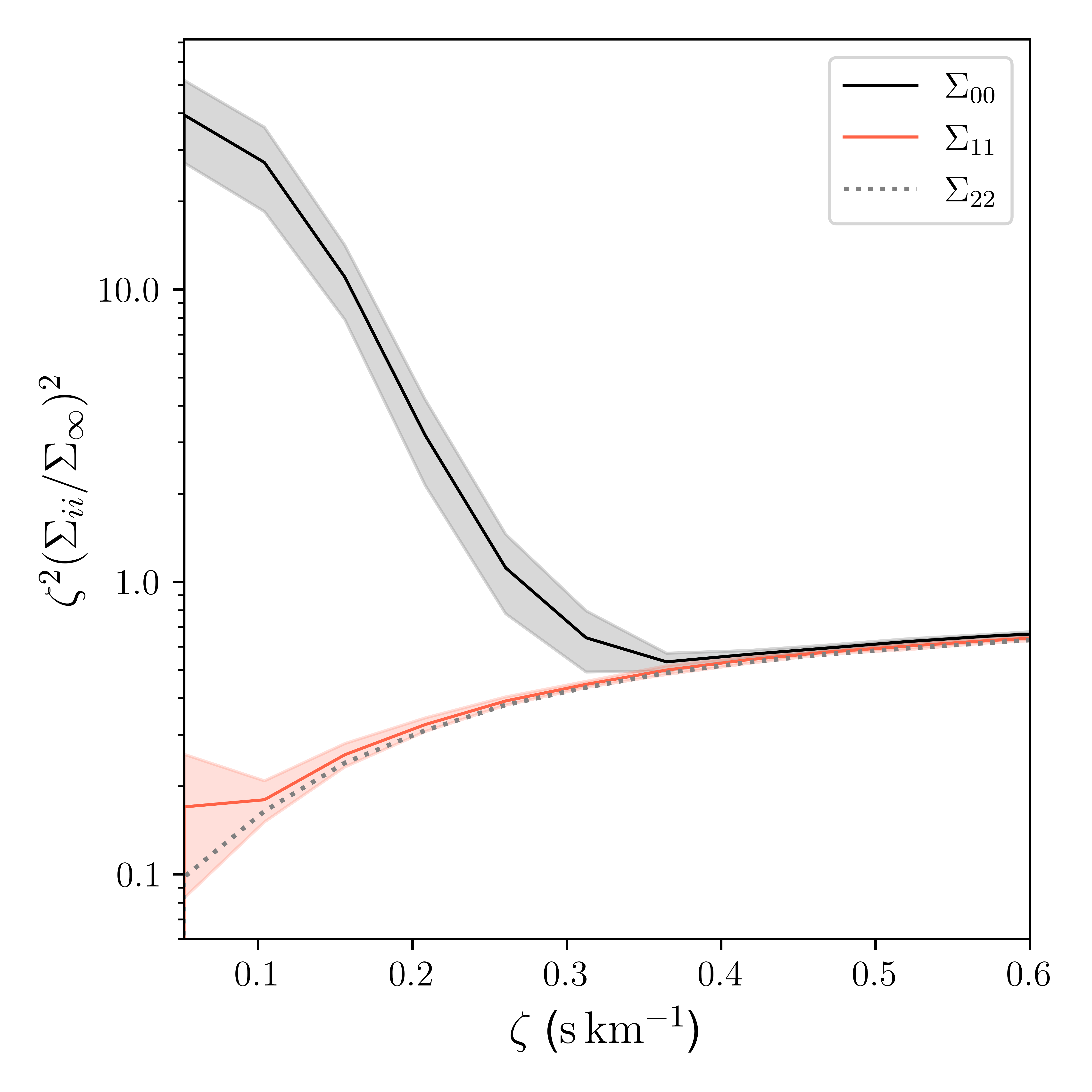} %
        \caption{HD~34411 Fisher information estimate for the first, second and third singular values versus 
        $\zeta$. Solid lines represent the median value of $\Sigma_{00}$ and $\Sigma_{11}$ from all orders used in the analysis, with colored band represents the one-sigma interval. A gray dotted line represents $\Sigma_{22}$ for reference. }
    \label{fig: HD 34411 singular values}
\end{figure}

The variance estimate in each order, presented in Figure~\ref{fig: HD 34411 order var}, shows that some orders contribute more information than others, possibly due to varying flux levels, differences in calibration accuracy, or imperfect telluric correction. Velocity is then extracted from each order as described above, and the estimates of all orders are then combined using inverse variance weights.
The Fisher information per-$\zeta$, shown in Figure~\ref{fig: HD 34411 singular values}, suggests that high-order-terms, and therefore their leakage to the zeroth order, are on average smaller than the noise level. We considered inverse wavelengths with $\sigma_{_\zeta}^{-1} > 10$ and did not account for leakage from the first order term.

\begin{figure}
        \centering         
        \includegraphics[width=0.8\columnwidth]{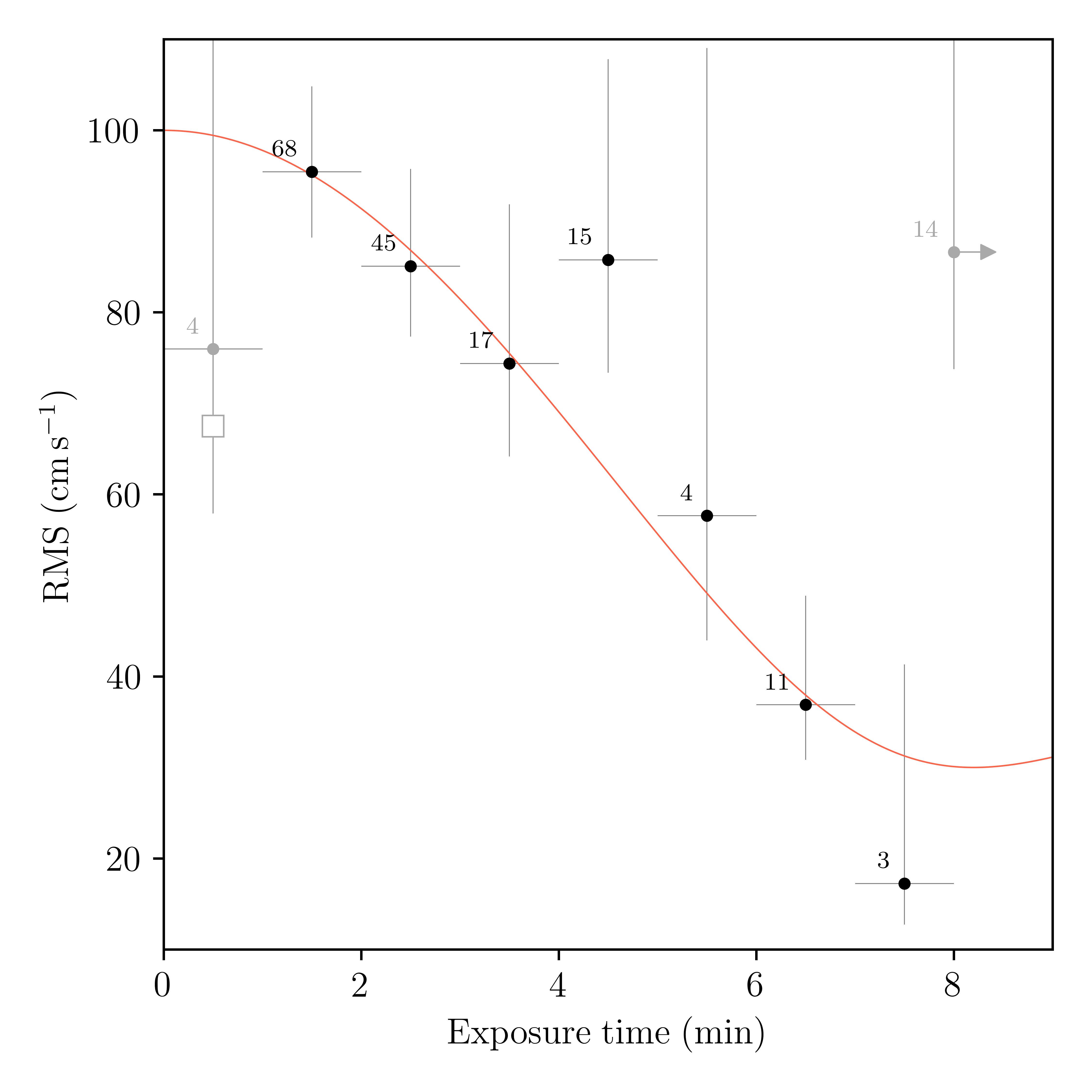} %
        \caption{HD~34411 radial velocity RMS at different exposure time bins. The error bars represent the RMS 1-$\sigma$ confidence interval and exposure time bin width. The number of points in each bin is presented alongside each measurement. For the sub-minute exposure, the black point represents only 4 non-consecutive measurements, where the RMS of all 11 measurements in this bin is shown as an empty square. The red curve presents the expected RMS trend due to p-mode averaging. The two rightmost outliers are assumed to be due to observing conditions.}
    \label{fig: HD 34411 RMS vs expT}
\end{figure}

As mentioned above, most observations were made with exposure times ($T_{\rm exp}$) shorter than the $8.2$~min p-mode cycle, which is expected to manifest as such additional scatter. To test this possibility, we subtract the injected sine-wave Doppler shift and calculate the root mean square (RMS) of the derived velocities for 1-minute exposure time bins, which as shown in Figure~\ref{fig: HD 34411 RMS vs expT}. We use Bessel's corrected estimate for the RMS, and calculate its corresponding 1-$\sigma$ confidence interval assuming it follows a $\chi^2_{n-1}$ distribution. The number of points in each bin is shown next to each point, with the exception of the leftmost bin. There are 11 points in this sub-minute bin, but only 4 of which are not consecutive. Since the sampling is short compared to the p-mode modulation, we consider the non-consecutive measurement to estimate the RMS in this bin.

Figure \ref{fig: HD 34411 RMS vs expT} shows the RMS decreasing as exposure times increase from 1 to 8 minutes, reaching about 30 cm s$^{-1}$ near the stellar p-mode period. This trend matches the expectation from p-mode averaging. A simple model of the residual velocity scatter, calibrated to give 100 cm s$^{-1}$ at one-minute exposures, is
\begin{equation}
    {\rm RMS}^2 \approx 9100\times\frac{\sin^2(\tau)}{\tau^2} + 900 \quad {\rm cm^2~s^{-2}},
\end{equation}
where $\tau \equiv \pi \, T_{\rm exp}\,/\,(8.2~{\rm min})$ and the second term represents the 30~cm~s$^{-1}$ noise floor. Most bins follow this trend, but the two longest exposure bins exceeds our expectations by a factor of ${\sim}\,2$. This is possibly due to sub-optimal observing conditions during those measurements. 

\begin{figure*}
        \centering           \includegraphics[width=1\textwidth,trim={2.5cm 0cm 6cm  0cm},clip]{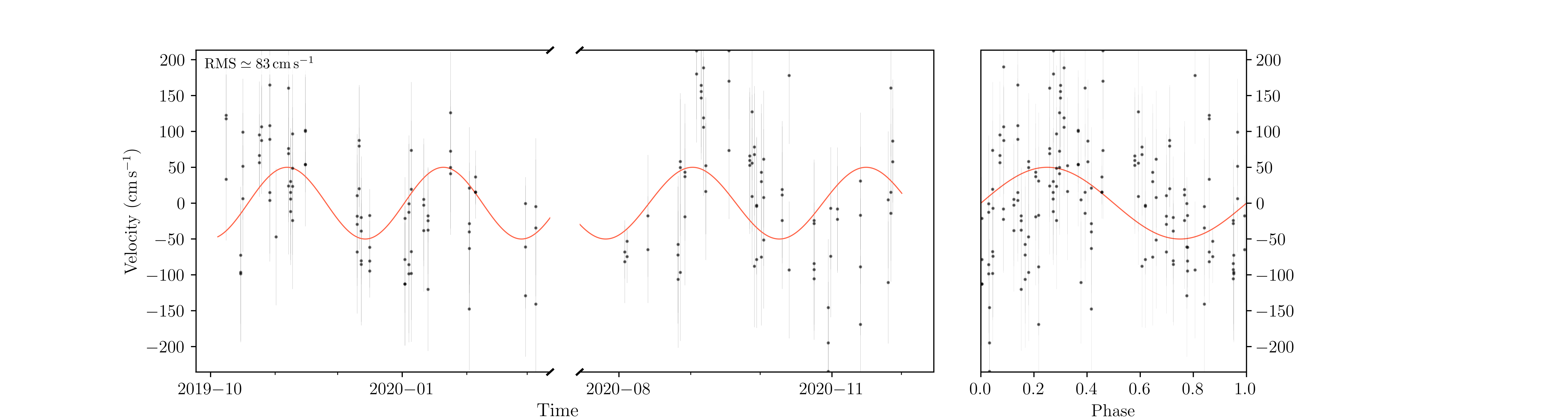} %
        \caption{\textit{Left panel}---The velocities extracted for the HD 34411 spectroscopic time series. The horizontal axis presents the month and year in which the data were obtained. The dash signs in the middle of the panel account for the gap in the observing campaign. The injected signal is shown as a solid red line. \textit{Right panel}---same as the left panel, but phase-folded with the injected period of 75 days. }
    \label{fig: HD 34411 rv}
\end{figure*}

Finally, we recover the semi-amplitude of the injected signal. To do so, we fit the radial velocity semi-amplitude, assuming the orbital period is known. We use a subset of 163 measurements with exposure times of $1-8$~minutes and recalibrate the measurement uncertainty to the estimated error on each exposure bin. This process yields
\begin{equation}
    \hat{v} = (47.3 \pm 6.8) \times\sin\left[\frac{2\pi(t-t_0)}{75~{\rm day}} \right] +  (-96.5 \pm 4.8) \quad {\rm cm\,s^{-1}}
\end{equation}
with reduced $\chi^2\simeq1$ (as expected, given that the scatter was derived assuming the solution is known). The inferred semi-amplitude is consistent with the injected signal. A non-zero bias term is required, as the radial velocity estimates are determined relative to a reference kernel. The velocities used in the fit are presented in Figure~\ref{fig: HD 34411 rv}.

\begin{figure}
    \centering
    \includegraphics[width=0.8\columnwidth]{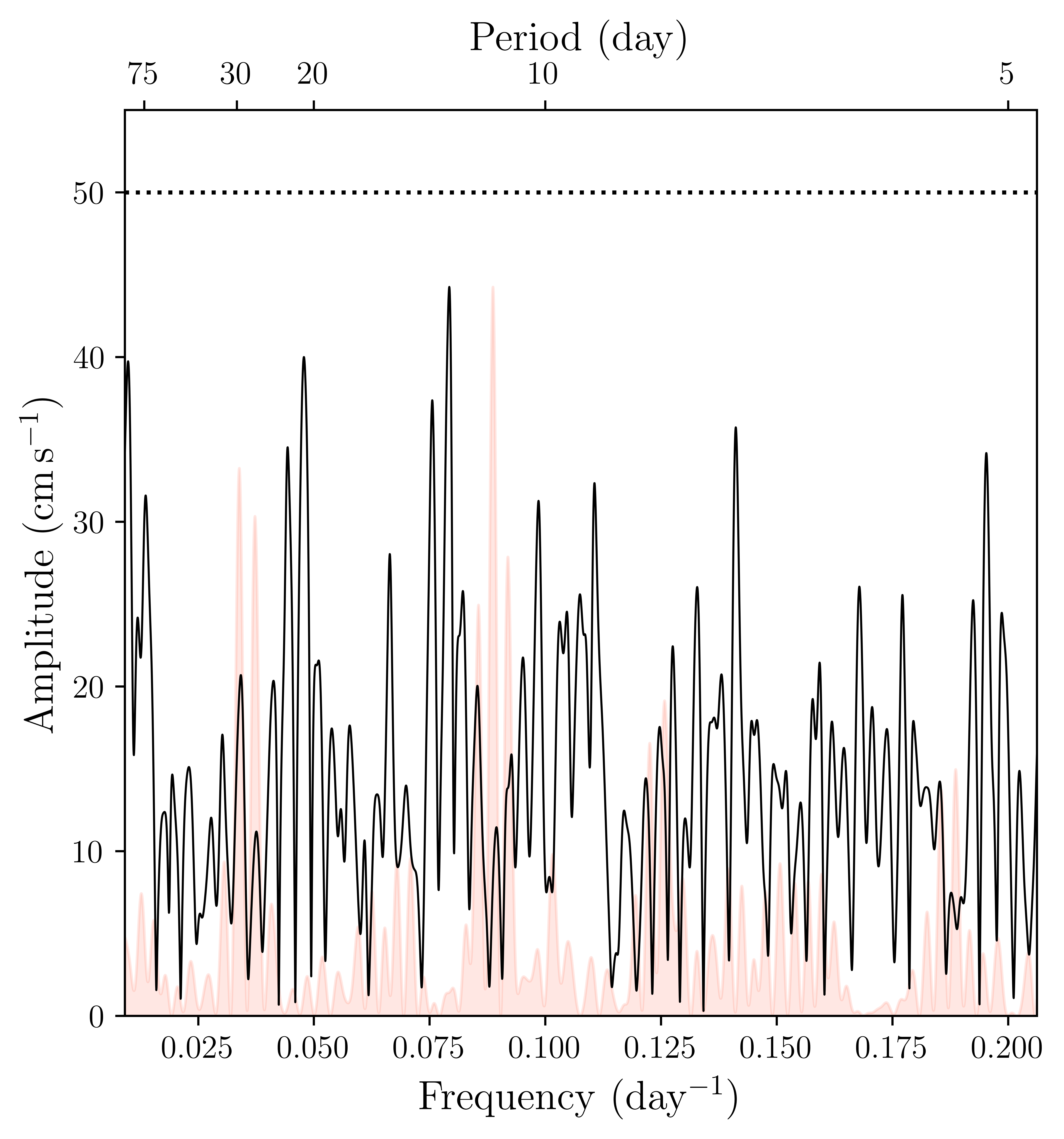}
    \caption{A periodogram of the velocities, after removing the 75-day injected signal. The uncertainties used in the calculation are as shown in Figure~\ref{fig: HD 34411 rv}. The solid black line represents the amplitude of a fitted harmonic model at each frequency. The red region represents the scaled window function of the time series. }
    \label{fig: HD 34411 spectrogram}
\end{figure}

Figure~\ref{fig: HD 34411 spectrogram} shows the periodogram of the derived velocities after subtracting the injected 75-day signal. The solid black curve marks the best-fitting harmonic amplitude at each frequency, while the red region traces the sampling-induced window function. All fitted amplitudes remain below about 40~cm\,s\textsuperscript{-1}, with 95\% of the frequencies yielding amplitudes under 30~cm\,s\textsuperscript{-1}. The apparent residual scatter is fully explained by p-mode modulation, finite exposure-time averaging, and the sampling window. These results confirm that, under favorable conditions and with appropriate integration times, EXPRES attains its nominal precision.


\subsection{Test case II: HD 10700}
\label{sec: tau cet}
HD 10700 ($\tau$~ Ceti) is a nearby G8V star at a distance of ${\sim}\,3.7$ pc. It has an apparent $V$-band magnitude of ${\sim}\,3.5$, an effective temperature of ${\sim}\,5300$ K, and a sub-solar metallicity of about $-0.5$ dex. Its low chromospheric activity and slow rotation are consistent with an estimated age of $8-12$ Gyr \citep{Tang2011, zhao22}.

The dataset includes 174 measurements taken over a 15.5-month baseline, with typical exposures of $1-2$ minutes.  Two intensive sub-campaigns, on 2019 August 10\textsuperscript{th} and October 25\textsuperscript{th}, provide sequences of ${\sim}\,20$ consecutive observations each. The selected wavelength range, sampling, and segment lengths are as before (see test cast~I above). In this case, we did not inject a synthetic signal to the spectroscopic dataset. 

\begin{figure*}
        \centering         
        \includegraphics[width=0.975\textwidth,trim={2.5cm 0 2.5cm 0},clip]{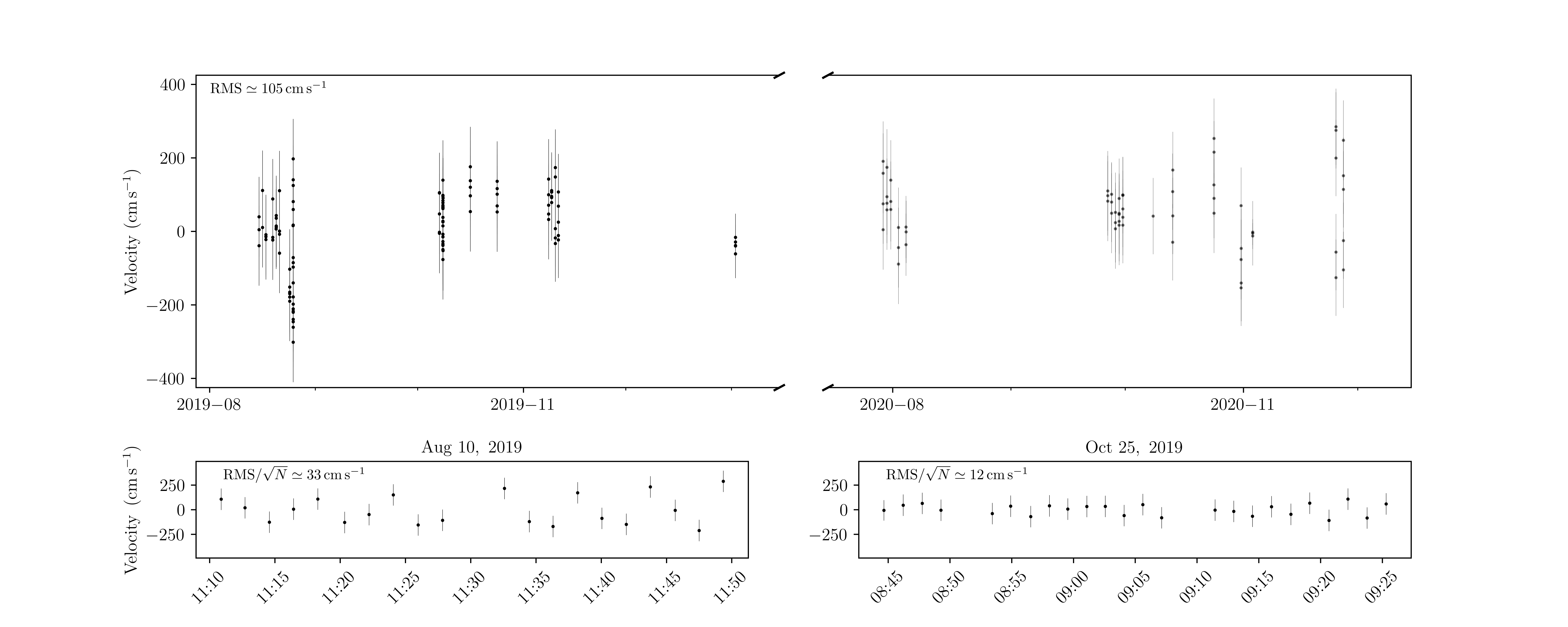} %
        \caption{\textit{Top panel}---The velocities extracted for the $\tau$-Ceti spectroscopic time series. The horizontal axis presents the month and year in which the data were obtained. The dash signs in the middle of the panel account for the gap in the observing campaign. \textit{Bottom panels}---a zoomed-in view of the high sampling cadence nights. The velocities on the left panel, taken on the night of August 10\textsuperscript{th} 2019, show enhanced variability with periodicity of 5--5.5 minutes, consistent with what is expected of pressure modes. }
    \label{fig: tau cet rv}
\end{figure*}

The derived velocities are shown in Figure~\ref{fig: tau cet rv}.
The figure shows the full $\tau$~Ceti radial velocity time series (top), with an overall RMS of ${\sim}\,100$~cm\,s\textsuperscript{-1}. Each black point marks an individual exposure; error bars reflect formal uncertainties. Two high-cadence nights are shown below. On August 10\textsuperscript{th}, 2019 (bottom left), the intra-night RMS reaches ${\sim}\,150$~cm\,s\textsuperscript{-1}, with a clear ${\sim}\,5$~min periodicity. On October 25\textsuperscript{th} (bottom right), the RMS is lower at  ${\sim}\,60$~cm\,s\textsuperscript{-1}. 
The two nights likely reflect stochastic changes in p-mode amplitudes, consistent with expectations for quiet G-type stars, and match those reported for the Sun \citep{Phillips2016, Lanza2016}. They likely arise from stochastic changes in the p-mode envelope, which combines the power of all modes near $\nu_{\rm max}$ and can vary in coherence and strength over days \citep[e.g.,][]{Kjeldsen1995}.

While the two well-sampled nights shown in the bottom panels of Figure~\ref{fig: tau cet rv} display clear differences in p-mode amplitude, they show no measurable zero-point drift. The inferred velocities of $\tau$~Ceti therefore demonstrate both the strong imprint of stellar pulsations and the stability of the underlying velocity series. This suggests that precision can be further improved through optimized integration times, consecutive sampling, and dedicated modeling of p-mode–induced variability.

\section{Summary and Conclusions}\label{sec: summary}

We presented a generalized formalism for radial velocity extraction based on the STFT. By decomposing a spectroscopic time series into principal spectra and time-dependent kernels, the method enables simultaneous recovery of the underlying spectral components and the radial velocity shifts, even when the principal spectra are unknown. The Doppler signal is encoded as coherent phase differences across epochs, allowing a statistically tractable and linearized solution in the high-SNR regime. Validation on synthetic datasets demonstrates robust performance when applied to simulated data, under both Gaussian noise and realistic stellar variability.

The STFT–SVD framework recovers coherent radial velocity signals at the few–tens of cm\,s$^{-1}$ level using existing high-resolution spectrographs. Applied to EXPRES observations of HD~34411 and $\tau$~Ceti, it achieves sub–m\,s$^{-1}$ precision and, for the available sampling and integration times, reaches the instrumental precision floor while capturing stochastic variability consistent with solar-like p-mode oscillations. These results demonstrate that the factorisation formalism can separate genuine Doppler shifts from intrinsic spectral variability, providing a practical route toward extreme-precision radial velocity extraction.

The method is not intrinsically limited to Sun-like stars or to visible wavelengths. Its applicability is governed by whether the spectral variability can be linearized into a small number of components, with sufficient phase coverage to separate Doppler shifts from higher-order terms. Faster rotation acts as a low-pass filter in Fourier space, reducing the number of informative inverse-wavelength modes. More complex or information-rich spectra, such as near-infrared observations of M dwarfs, require greater sampling to avoid degeneracies. These constraints are generic to Doppler spectroscopy and are not specific to this framework.

Still, the technique was benchmarked on a limited set of targets and observing conditions, which is insufficient to fully characterise its limitations. Sampling cadence and integration times were fixed by the original observing programmes, inherently constraining the inference. We did not explore alternative sampling strategies, exposure distributions, or the impact of correlated noise. The SNR required to robustly detect chromatic or higher-order terms, and their sensitivity to instrumental systematics, also remains unconstrained. Nevertheless, the results provide a clear proof of concept: under realistic conditions, the factorisation approach reaches sub–m\,s$^{-1}$ precision and offers a viable, data-driven framework for modelling spectral variability.

Extending the method to highly active stars requires sufficient phase coverage of activity-driven spectral variability and, ideally, independent constraints on the temporal structure of the activity signal. A further complication arises from the treatment of spectral orders: unlike full-spectrum simulations, per-order decompositions do not guarantee consistent ordering or relative weighting of higher-order terms across orders. A systematic investigation of these effects, and their impact on active stars, is therefore deferred to future work. In parallel, we are developing a rigorous, observation-driven framework to mitigate p-mode modulation through optimized observing strategies and data-averaging schemes.

Beyond exoplanet detection, the formalism opens a path to broader astrophysical applications. Its ability to reconstruct spatially varying contributions to the stellar spectrum suggests future use in Doppler imaging and tomography. In particular, it may enable obliquity measurements during planetary transits without requiring strong assumptions about the stellar rotation profile or surface properties. By retrieving the shapes of the principal spectra and tracking the behaviour of higher-order kernels, the decomposition provides a framework for studying stellar surface structure, centre-to-limb variations, and time-domain spectral evolution in a data-driven manner. Future work will extend the method to additional targets and applications, advancing its role in empirical stellar characterisation and precision spectroscopy.

\section*{Acknowledgements}
We thank Lily L.~Zhao for valuable comments and for sharing the EXPRES data, and Yinan~Zhao and Xavier~Dumusque for providing the SOAP simulations. We are also grateful to Sagi~Ben-Ami, Eran~O.~Ofek, and Avishay~Gal-Yam for their advice and support.  We thank the anonymous referee for the helpful advice. 

This research is supported by the Israeli Council for Higher Education (CHE) via the Weizmann Data Science Research Center, and by a research grant from the Estate of Harry Schutzman. SS acknowledge support from the European Research Council for the ERC Advanced Grant [101054731]. SS was supported by the Benoziyo Prize Postdoctoral Fellowship at the Weizmann Institute of Science, where the majority of this research was carried out. BZ is supported by a research grant from the Willner Family Leadership Institute for the Weizmann Institute of Science.

These results made use of the Lowell Discovery Telescope at Lowell Observatory. Lowell is a private, non-profit institution dedicated to astrophysical research and public appreciation of astronomy and operates the LDT in partnership with Boston University, the University of Maryland, the University of Toledo, Northern Arizona University and Yale University. The EXPRES team acknowledges support for the design and construction of EXPRES from NSF MRI-1429365, NSF ATI-1509436, and Yale University. DAF gratefully acknowledges support to carry out this research from NSF 2009528, NSF 1616086, NSF AST-2009528, the Heising-Simons Foundation, and an anonymous donor in the Yale alumni community. 

SS dedicates this work to the memory of Gil Ezra, Amit Levi and Itamar Tal, and to their families.
\section*{Data Availability}
data and code underlying the simulations are available via an online repository (10.5281/zenodo.18628991). The EXPRES data underlying this article were provided by Extreme Stellar Signals Project (ESSP) by permission. Data will be shared on request to the corresponding author with permission of ESSP coordinators. 
 



\bibliographystyle{mnras}
\bibliography{main} 




\appendix

\section{Matrix form}\label{app: matrix form}

The real parts of the complex phase and kernel ratio remain close to unity in the limit of very small complex phases. Moreover, in this limit, the imaginary part of the exponent is approximately equal to its argument. Consequently,
\begin{equation}
    \left|\exp(-i2\pi\zeta v ) - \mathcal{R}^{(0)}_{_\zeta}\right|^2 \simeq \left[2\pi\zeta v + \Im(\mathcal{R}^{(0)}_{_{\zeta, i}})\right]^2,
\end{equation}
and log-likelihood can be linearized in the velocities of different epochs, $v_{_i}$,
As a result, the likelihood takes the form 
\begin{equation}
    \log\mathcal{L} \simeq -\sum_{_{\zeta i}}\sigma_{_\zeta}^{-2}\left[2\pi\zeta v_{_i} + \Im(\mathcal{R}^{(0)}_{_{\zeta, i}}) \right]^2 + \text{const}.
\end{equation}
The log-likelihood above is linearized in the velocities of different epochs, $v_{_i}$.

This linearized likelihood can be maximized using a standard approach with the appropriate design matrix. For each inverse wavelength, we define a matrix block of size $N_{\rm obs} \times N_{\rm obs}$, where $N_{\rm obs}$ is the number of spectra in the time series. The block corresponding to the $k$\textsuperscript{th} inverse wavelength is given by
\begin{equation}\label{eq: design matrix}
    \mathbf{X}_{_k}[i,j] =
    \begin{cases}
        2\pi\zeta_k & \quad \text{if } \,\, i=j,\\
        0           & \quad \text{otherwise}.
    \end{cases}
\end{equation}
These blocks are stacked to form the spectroscopic time series design matrix. We aim to find a vector of velocities, $\mathbf{v}$, that solves 
\begin{equation}\label{eq: rv equation system}
    \begin{bmatrix}
        \mathbf{X}_{_1} \\[3pt]
        \vdots \\[3pt]
        \mathbf{X}_{N_\zeta}
    \end{bmatrix}
    \,
    \mathbf{v}
    =
  -\Im  \begin{bmatrix}
        \boldsymbol{\mathcal{R}}^{(0)}_{_{1}} \\
        \vdots \\
        \boldsymbol{\mathcal{R}}^{(0)}_{_{N_\zeta}}
    \end{bmatrix}
    + \boldsymbol{\epsilon},
\end{equation}
where $\boldsymbol{\mathcal{R}}^{(0)}_{_{k}}$ represents a vector of the complex ratio of the $k$\textsuperscript{th} inverse wavelength of all measurements, $N_{_\zeta}$ is the number of inverse wavelengths used in the analysis, and $\boldsymbol{\epsilon}$ is a vector of Gaussian errors defined per inverse wavelength, as described in equation~(\ref{eq: R complex normal}). From the dimensions of the design matrix, it follows that the data set provides $N_{\rm obs} \times N_\zeta$ constraints for $N_{\rm obs}$ parameters. The least-squares solution to equation~(\ref{eq: rv equation system}) yields the radial velocities.

\section{Perturbed principal kernels}
\label{app: perturbed SVD}
This appendix shows a first-order perturbation analysis to quantify how additive noise distorts the singular values and vectors. The derivation starts from the spectral decomposition of the corresponding Gram matrix and retains only the leading-order terms in the noise matrix, in a Rayleigh-Schr\"{o}dinger perturbation approach. For sharper, non-asymptotic bounds, see \citet{ORourke2024} and references therein.

Let ${\bf A}\in\mathbb{C}^{m\times n}$ be the ideal, noise-free data matrix, whose singular-value decomposition is  
\begin{equation}
    {\bf A} \equiv \sum_{i}\sigma_{i}\,{\bf u}_i{\bf v}_i^{\dagger},
    \qquad \text{where}\,\,
    \sigma_0\ge\sigma_1\ge\dots .
\end{equation}
The measured matrix includes a noise term, expressed as a matrix of random entries, ${\bf E}$. Hence,
\begin{equation}
    \tilde{\bf A} = {\bf A}+{\bf E}
                   = \sum_{i}\tilde{\sigma}_i\,\tilde{\bf u}_i\tilde{\bf v}_i^{\dagger},
\end{equation}
where the term on the left represents the decomposition of the measured matrix.
We aim to express the observed left-side singular vectors as a series expansion of the ideal unknown ones. For this purpose, we use the conjugate transposed Gram matrix, ${\bf M}\;\equiv\;{\bf A}{\bf A}^{\dagger}$. As before, the observed matrix differs from the ideal, noise-free one. It is described to a first order as 
\begin{equation}
    \tilde{\bf M}\equiv \tilde{\bf A}\tilde{\bf A}^{\dagger}= {\bf M} + \epsilon\,{\bf W} + {\cal O}(\epsilon^2),
\end{equation}
where $\epsilon$ is a small expansion parameter and ${\bf W}$ is the additive matrix that represents the first-order deviation due to the noise term. This is a somewhat loose definition. Practically, the expansion parameter $\epsilon$ can be thought of, for example, as the maximal standard deviation of all random entries of ${\bf E}$.

For a given left singular vector, we write the perturbative ansatz  
\begin{equation}
    \tilde{\bf u}_k = {\bf u}_k + \epsilon\,{\bf w}_k + {\cal O}(\epsilon^2),
\end{equation}
where $\langle{\bf u}_k,{\bf w}_k\rangle = 0$.
Considering that $\tilde{\bf M}\tilde{\bf u}_k = \tilde{\sigma}_k^{\,2}\tilde{\bf u}_k$, we insert the equation above and retain terms of order $\epsilon$. Rearranging the first-order perturbation terms, we get a relation 
\begin{equation}
    \bigl({\bf M}-\sigma_k^{2}{\bf I}\bigr){\bf w}_k
    = -\bigl({\bf W} - \lambda_k {\bf I}\bigr) {\bf u}_k,
\end{equation}
where $\lambda_k$ denotes the first-order correction to $\sigma_k^{2}$.
Projecting onto an unperturbed eigenvector ${\bf u}_j$ with $j\neq k$ yields
\begin{equation}
    \bigl(\sigma_j^{2}-\sigma_k^{2}\bigr)\,
    \langle{\bf u}_j,{\bf w}_k\rangle
    = -\langle{\bf u}_j,{\bf W}{\bf u}_k\rangle .
\end{equation}
Therefore, to first order in $\epsilon$, the inner product between a perturbed and unperturbed singular vector is 
\begin{equation}
\langle{\bf u}_j,\tilde{\bf u}_k\rangle \equiv \epsilon_{jk} =
\begin{cases}
    \epsilon\frac{{\langle{\bf u}_j,{\bf W}{\bf u}_k\rangle}}{{\sigma^2_k - \sigma^2_j}} & \text{if} \,\,\, j\neq k \,\,\, \text{and}\\[8pt]
    1  & \text{else}.
\end{cases}
\label{eq:leakage}
\end{equation}
We note that we did not account for the random phase induced by the SVD in the analysis above. This random phase will not change the magnitude of the contribution of each term in the series. Instead, it can be considered an unknown complex phase induced on the coupling coefficient. For simplicity, we will neglect the difference in the random complex phase in the discussion below and will also treat the signs of the coupling coefficients loosely.

Equation~\eqref{eq:leakage} shows that the closer two singular values are to one another, the more substantial the leakage between their corresponding singular vectors. For instance, the difference between the perturbed and unperturbed second left singular vectors, up to a random complex phase, is
\begin{equation}
    \tilde{\bf u}_0 - {\bf u}_0 \simeq {\epsilon_{10}} 
    \,{\bf u}_1 + {\epsilon_{20}}\, {\bf u}_2 +  \dots  +{\epsilon_{m0}}\,{\bf u}_m .
\end{equation}
The perturbation of the matrix causes leakage from all singular vectors. However, this case's dominant term is probably ${\bf u}_0$, because $\sigma_1$ is the closest neighbour of $\sigma_0$. If there are only two non-negligible eigen vectors, the terms of higher order will effectively behave as additive noise, and the expression is reduced to
\begin{equation}
    \tilde{\bf u}_0  \simeq {\bf u}_0 + \epsilon_{10}\,{\bf u}_1 + \text{noise}.
\end{equation}
Using similar considerations, the singular vector of the following order will probably attain the form
\begin{equation}
    \tilde{\bf u}_1  \simeq {\bf u}_1 + \epsilon_{20}\,{\bf u}_2 + \text{noise}.
\end{equation}
This is because, in the cases considered in our worked examples, $\sigma^2_1 - \sigma^2_2 \ll \sigma^2_0 - \sigma^2_1$. As a result, ${\bf u}_2$ is the leading term in the expansion of $\tilde{\bf u}_1$.

To estimate the scale of $\epsilon_{01}$, we note that the inner product of two independent unit-norm Gaussian vectors has variance $\sim n^{-1}$, where $n$ is the number of measurements. In our case,
\begin{equation}
|\epsilon_{01}| \lesssim \frac{\epsilon}{\sqrt{n}} \cdot \frac{1}{\sigma_0 + \sigma_1} \cdot \frac{1}{\sigma_0 - \sigma_1} \sim \frac{\epsilon}{\sqrt{n}} \frac{1}{\sigma_0^2}.
\end{equation}
The $1/\sqrt{n}$ factor reflects averaging over independent samples. The second term scales with the amplitude of the perturbation, while the last term represents the impact of the spectral gap: as $\sigma_1$ approaches $\sigma_0$, the decomposition becomes increasingly unstable.
For the cases considered here, $\sigma_0$ dominates all other singular values.

The perturbation considered in this equation does not represent white noise (see main text), but by structured leakage from the first-order mode. Since we are interested in cross-terms between the zeroth and first principal vectors, we assume the perturbations as  $\epsilon \propto \sigma_0 \times \sigma_1$, obtaining a simplified bound,
\begin{equation}\label{eq: leak bound}
|\epsilon_{01}| \propto \frac{1}{\sqrt{n}} \cdot \frac{\sigma_1}{\sigma_0}.
\end{equation}

\end{document}